\documentclass[journal,10pt]{IEEEtran}
\usepackage{amssymb}
\usepackage{amsmath}
\usepackage{cite}
\usepackage{url}
\usepackage{xcolor}
\usepackage{cite,graphicx,amsmath,amssymb}
\usepackage{graphicx}
\usepackage{subfigure}
\usepackage{citesort}
\usepackage{fancyhdr}
\usepackage{mdwmath}
\usepackage{mdwtab}
\usepackage{caption}
\usepackage{amsthm}
\usepackage{setspace}
\usepackage{multirow}
\usepackage{epstopdf}
\usepackage[justification=centering]{caption}

\newtheorem{theorem}{Theorem}

\newtheorem{lemma}{Lemma}

\newtheorem{corollary}{Corollary}

\newtheorem{proposition}{Proposition}

\usepackage{algorithm}
\usepackage{algorithmic}
\usepackage{cases}

\makeatletter
\def\ScaleIfNeeded{%
	\ifdim\Gin@nat@width>\linewidth \linewidth \else \Gin@nat@width
	\fi } \makeatother

\begin{document}
	
	\title{Joint Design for Simultaneously Transmitting And Reflecting (STAR) RIS Assisted NOMA Systems}
	\author{
		 Jiakuo~Zuo,
		 Yuanwei~Liu,~\IEEEmembership{Senior Member,~IEEE,}
		 Zhiguo~Ding,~\IEEEmembership{Fellow, IEEE,}
		 Lingyang~Song,~\IEEEmembership{Fellow, IEEE,}
		 and H. Vincent Poor,~\IEEEmembership{Life Fellow, IEEE}
	   \thanks{Part of this work was presented at the IEEE Global Communications Conference, Madrid, Spain, 7-11 December 2021~\cite{zuojiakuo}.}
		\thanks{J. Zuo is with the School of Internet of Things, Nanjing University of Posts and Telecommunications, Nanjing 210003, China (e-mail: zuojiakuo@njupt.edu.cn).}
		\thanks{Y. Liu is with the School of Electronic Engineering and Computer Science, Queen Mary University of London, London E1 4NS, U.K.  (e-mail: yuanwei.liu@qmul.ac.uk).}
		 \thanks{Z. Ding is with the School of Electrical and Electronic Engineering, University of Manchester, Manchester, U.K. (email: zhiguo.ding@manchester.ac.uk).}
	 	\thanks{ L. Song is with Department of Electronics, Peking University, Beijing 100871 China (email: lingyang.song@pku.edu.cn).}
	 	\thanks{H. V. Poor is with the Department of Electrical and Computer Engineering, Princeton University, Princeton, NJ 08544, USA. (email: poor@princeton.edu).}				
	}
	\maketitle
	\vspace{-1.5cm}
 \begin{abstract}
	Different from traditional reflection-only reconfigurable intelligent surfaces (RISs), simultaneously transmitting and reflecting RISs (STAR-RISs) represent a novel technology, which  extends the \textit{half-space} coverage to \textit{full-space} coverage by simultaneously transmitting and reflecting incident signals. STAR-RISs provide new degrees-of-freedom (DoF) for manipulating signal propagation. Motivated by the above, a novel STAR-RIS assisted non-orthogonal multiple access (NOMA) (STAR-RIS-NOMA) system is proposed in this paper. Our objective is to maximize the achievable sum rate by jointly optimizing the decoding order, power allocation coefficients, active beamforming, and transmission and reflection beamforming. However, the formulated problem is non-convex with intricately coupled variables. To tackle this challenge, a suboptimal two-layer iterative algorithm is proposed. Specifically, in the inner-layer iteration, for a given decoding order, the power allocation coefficients, active beamforming, transmission and reflection beamforming are optimized alternatingly. For the outer-layer iteration, the decoding order of NOMA users in each cluster is updated with the solutions obtained from the inner-layer iteration. Moreover, an efficient decoding order determination scheme is proposed based on the equivalent-combined channel gains. Simulation results are provided to demonstrate that the
proposed \textcolor[rgb]{0.00,0.00,0.00}{STAR-RIS-NOMA} system, aided by our proposed algorithm, outperforms conventional RIS-NOMA and RIS assisted orthogonal multiple access (RIS-OMA) systems.
 \end{abstract}

 
 \section{Introduction}
 	Reconfigurable intelligent surfaces (RISs) are promising candidates for improving the performance of future sixth-generation (6G) wireless communication networks~\cite{9502643,Wu2020TowardsSA}. RISs are planar arrays consisting of large numbers of low-cost reconfigurable passive elements. By properly adjusting the amplitude and phase response of these elements, the propagation of incident wireless signals can be reconfigured. It is worth noting that RISs mainly constitute passive devices without the need for active radio frequency chains. Therefore, RISs are economically and environmentally friendly, and can be densely deployed in wireless networks with low cost and low energy consumption~\cite{Renzo2020SmartRE}. Due to the aforementioned attractive characteristics, RISs have received considerable attention from both industry and the research community. 
 	
 	Much of the existing research on RISs assume that 
 	they can only reflect incident signals, which requires that the transmitter and receiver must be located on the same side of the RIS. Thus, this geographical constraint, i.e., \textit{half-space }coverage, limits the flexibility of RIS deployment. To overcome this limitation, recently, a novel type of RIS, termed simultaneous transmitting and reflecting RISs (STAR-RISs)~\cite{STAR} or intelligent omni-surfaces (IOSs)~\cite{Zhang2021IntelligentRM}, has been proposed. Different from traditional reflection-only
 	RISs, STAR-RISs can simultaneously transmit and reflect the
 	incident signals, which leads to a \textit{full-space} coverage. To support simultaneous transmission\footnote{\textcolor[rgb]{0.00,0.00,0.00}{In physics fields, 'transmission' is commonly known as 'refraction'~\cite{LaSpada2018CurvilinearMF,Pfeiffer2013MetamaterialHS}. In this paper, we use 'transmission/transmitted'.}} and reflection, the elements of
 	STAR-RISs need to support both electric and magnetic currents~\cite{STAR}. As a result, the transmitted and reflected signals can be reconfigured by a STAR-RIS element via its corresponding transmission and reflection coefficients, which introduces additional degrees-of-freedoms (DoF) to control the signal propagation.  
 	
 	On the other hand, as a promising technique for enhancing spectral efficiency and supporting massive connectivity, non-orthogonal multiple access (NOMA) has also received significant attention~\cite{7842433}. NOMA outperforms conventional orthogonal multiple access (OMA) techniques by simultaneously sharing the
 	communication resources between all users via the power or
 	code domain~\cite{liu2018non}. \textcolor[rgb]{0.00,0.00,0.00}{Inspired by the advantages of RIS and NOMA, the combination of the two techniques can provide the following performance gains, which are summarized as follows~\cite{9424177,9241881}: 1) In conventional NOMA networks, the decoding order is determined by the channel conditions, which are determined by the environment. However, the channel quality can be enhanced or degrade by adjusting the RIS coefficients in RIS assisted NOMA (RIS-NOMA) systems. This leads to the transition from '\textit{channel condition-based NOMA}' to '\textit{quality of service (QoS)-based NOMA}'. 2) Providing uniform signal coverage for conventional NOMA networks is difficult and serving users with poor channel conditions will reduce the system sum-rate. By introducing RISs, long-range communication is enabled for NOMA networks and more users can be served. As a result, the connectivity capacity will be enhanced by RIS-NOMA systems. 3) The reflection links of RISs can boost the channel gains to improve the network performance. Meanwhile, with dynamic power allocation methods, the users can achieve a similar data rates, which guarantees the fairness. 4) In conventional NOMA systems, to improve the data rate of a weak user, the power of the weak user must be increased and the other users' power must be decreased. Since more power are allocated to weak user and the system sum rate may be decreased, which leads to a low energy efficiency. Fortunately, RISs can improve the weak users' data rate without requiring extra energy. Thus, the energy efficiency can be improved by RIS-NOMA systems. }
 	
 	Motivated by the RIS-NOMA systems, the purpose of this paper is to investigate promising applications of the STAR-RIS technique in NOMA systems for further performance improvement.  
\subsection{Related Works}
\subsubsection{Studies on RIS Assisted NOMA Systems}
RIS-NOMA systems have been intensively investigated. For example, the single-input single-output (SISO) RIS-NOMA systems were studied in~\cite{9167258,zheng2020intelligent,9353406}, where the resource allocation (such as sub-channel assignment and power allocation) and phase shift optimization problems were jointly optimized. In~\cite{9380234,9234527,9240028}, the total transmit power minimization problem for multiple-input-single-output (MISO) RIS-NOMA systems was investigated by jointly optimizing the active beamforming at the base station (BS) and passive beamforming at the RIS. In~\cite{9197675}, the energy efficiency maximization problem for MISO RIS-NOMA was considered. The formulated non-convex problem was solved by the alternating optimization, semi-definite programming (SDP) and successive convex approximation (SCA) approaches. RIS assisted millimeter-Wave (mmWave) NOMA systems were considered in~\cite{9140006}, where the analog beamforming and/or digital beamforming, power allocation and passive beamforming were jointly optimized. The physical layer security of RIS-NOMA systems was introduced in~\cite{Zhang2021SecuringNN,9385957} and a problem of maximizing the minimum secrecy rate was formulated by jointly optimizing active beamforming and passive beamforming. 
In the above works, all the involved channel state information (CSI) was assumed to perfect; however in~\cite{Zhang2021RobustAS,9279247}, the scenario of imperfect CSI for RIS-NOMA system was considered.
\subsubsection{Studies on STAR-RIS Assisted Communication Systems}
  In~\cite{Xu2021SRARRISsSR}, the channel models for the near- and far-field regions of STAR-RISs were proposed and closed-form expressions for channel gains of users receiving the transmission and reflection signal were derived. In~\cite{9462949}, the fundamental coverage characterization of STAR-RIS assisted two-user communication networks was investigated. The sum coverage range maximization problems were formulated for both NOMA and OMA systems. The power consumption minimization problems for STAR-RIS assisted unicast and multicast systems were studied in~\cite{9570143}, where active and passive beamforming were jointly optimized for different operating protocols of STAR-RIS. The application of IOS in an indoor multi-user downlink communication system was studied in~\cite{9491943}, where a joint IOS analog beamforming and small base station digital beamforming optimization problem was formulated to maximize the sum-rate of the system. In~\cite{Zhang2020BeyondIR}, the optimization of phase shifts of an IOS was analyzed and a branch-and-bound based algorithm was proposed to design the IOS phase shifts in a finite set.
 \subsection{Motivation and Contributions}
The advantages of deploying STAR-RISs in wireless communications are as follows: 
\begin{enumerate}
	\item The communication coverage can be extended to full space by simultaneously transmitting and reflecting incident signals.
	\item STAR-RISs provide new DoF for system designs, which allows us to optimize both transmission and reflection coefficients.
\end{enumerate} 

 In order to exploit the full potential of STAR-RISs, the active beamforming/resource allocation at the BS, and the transmission and reflection coefficients at the STAR-RIS need to be jointly optimized. However, it is challenging to solve this joint optimization problem because of the highly coupled optimization variables. Therefore, how to jointly optimize these variables is a fundamental problem for STAR-RIS assisted wireless communication systems. \textcolor[rgb]{0.00,0.00,0.00}{On the other hand, by utilizing \textcolor[rgb]{0.00,0.00,0.00}{successive interference cancellation (SIC)}  technique, NOMA can partially remove the cochannel interference at the receivers. However, with the raising number of users, NOMA system will become complicated. Because the complexity of utilizing SIC technique will increase. One efficient way to reduce the complexity is to employ cluster-based NOMA techniques which partition users into different groups. However, new interference, namely, inter-cluster interference, is introduced by cluster-based NOMA techniques. To reduce the inter-cluster and the intra-cluster interference, power allocation and beamforming optimization techniques play important role.}  
 
 In light of the above background and to the best of our knowledge, the joint optimization design for the STAR-RIS assisted NOMA (STAR-RIS-NOMA) system has not been studied yet, which motivates this work.  \textcolor[rgb]{0.00,0.00,0.00}{The main contributions of this paper are summarized as follows}\footnote{\textcolor[rgb]{0.00,0.00,0.00}{In preliminary work~\cite{zuojiakuo}, we briefly introduced the STAR-RIS-NOMA system and the proposed algorithms. The conclusions are provided without proof. However, in this paper, the operation principle and signal model of the STAR-RIS are clarified. The formulated problem and proposed algorithm  are analyzed in details. Furthermore, we provide the proofs of the conclusions. More simulation results are conducted to evaluate the performance of the proposed system and algorithm.}}:
 \begin{enumerate}
 	\item We propose a downlink STAR-RIS-NOMA communication system, where a separate STAR-RIS assists the
 	communication from the BS to the clustered users, and the users are randomly distributed in the transmission and reflection spaces of the STAR-RIS.
 	Specifically, we jointly optimize the decoding order, power allocation coefficients, active beamforming, transmission and reflection beamforming for maximizing the achievable sum rate of all users, subject to minimum QoS requirements, SIC decoding conditions, total transmit power constraint, transmission and reflection coefficients constraints. However, the objective function is not jointly concave over the optimization variables, which are highly coupled. To tackle this challenging problem, we first determine the decoding order for the NOMA users in each cluster. Then, for a fixed decoding order, the power allocation coefficients, active beamforming, transmission and reflection beamforming are determined alternatingly.
 	\item We propose a novel scheme to determine the decoding order according to the equivalent-combined channel gains and we also prove that the SIC condition can be guaranteed under this decoding order. It is worth pointing out that the decoding order is determined by the vectors of active, transmission and reflection beamforming, but has no relationship with the power allocation coefficients.	
 	 \item For a given decoding order, we divide the original problem into three sub-problems, and solve them alternatingly. In particular, we derive an optimal power allocation strategy in closed-form for the power allocation coefficient optimization problem. We utilize the SCA and SDP methods to solve the active beamforming optimization problem. For the joint transmission and reflection beamforming optimization problem, we propose an efficient iterative algorithm by applying a sequential constraint relaxation algorithm. Finally, we develop a novel two-layer iterative algorithm for the STAR-RIS-NOMA system, where the outer-layer iteration updates the decoding order, and the inner-layer iteration updates the power allocation coefficients, active beamforming vectors, transmission and reflection beamforming vectors alternatingly.
 	  \item Our simulation results show: 1) the performance gain of the STAR-RIS-NOMA system can be significantly enhanced by increasing the number of RIS elements; 2) the proposed decoding order determination scheme can achieve near-optimal performance; and 3) by employing STAR-RISs, the proposed STAR-RIS-NOMA system outperforms traditional RIS-NOMA and RIS-OMA systems.
 \end{enumerate}
 \vspace{-0.3cm}
\subsection{Organization}
The rest of this paper is organized as follows. \textcolor[rgb]{0.00,0.00,0.00}{In Section II, a brief introduction to the physics principle and basic signal model of the STAR-RIS is provided.} In Section III, the system model and problem formulation for designing the STAR-RIS-NOMA system are presented. In Section IV, we propose a two-layer iterative algorithm to solve the original optimization problem. Numerical results are presented in Section V, which is followed by the conclusions in Section VI.

Notation: $\mathcal{C}^{M \times 1}$ denotes a complex vector of dimension \emph{M}, diag(\textbf{x}) denotes a diagonal matrix whose diagonal elements are the corresponding elements in the vector \textbf{x}. The $m$-th element of a vector $\textbf{x}$ is denoted by $\left[ \mathbf{x} \right] _m$ and the $(m,n)$-th element of a matrix $\textbf{X}$ is denoted by $\left[ \mathbf{X} \right] _{m,n}$. ${\textbf{x}}^{H}$ denotes the conjugate transpose of the vector \textbf{x}. Tr(\textbf{X}) and rank(\textbf{X}) denote the trace and rank of a matrix \textbf{X}, respectively. $\mathcal{C}\mathcal{N}\left( 0,\sigma ^2 \right) $ denotes a complex Gaussian distribution with zero mean and variance $\sigma ^2$. A summary of key variables is presented in Table~\ref{Notations} for convenience of the readers.
\begin{table*}[ht]
	\setlength{\belowcaptionskip}{-10pt}
	\caption{Summary of Key Variables}
	\label{Notations}
	\begin{center}
		\begin{tabular} {|c|p{9cm}<{\centering}|}
			\hline
			Variables      &   Descriptions \\
			\hline
			$s_m$, $t_m$,$r_m$   & The incident, transmitted and reflected signals on the $m$-th element of the STAR-RIS \\
			\hline			
			 $\beta _{m}^{t}\left( \beta _{m}^{r} \right) ,\theta _{m}^{t}\left( \theta _{m}^{r} \right)$
			 	&   The amplitude and phase shift of transmission (reflection) coefficients		\\
			\hline
			 \textcolor[rgb]{0.00,0.00,0.00}{ $\mathcal{M},  \mathcal{C} , \textcolor[rgb]{0.00,0.00,0.00}{ \mathcal{K} }, \textcolor[rgb]{0.00,0.00,0.00}{ \mathcal{K} }_c$ }       	& The set of total STAR-RIS elements, clusters, users, and users in cluster $c$ 			\\
			\hline
			  $\mathbf{u}_t\left( \mathbf{u}_r \right) ,\mathbf{\Theta }_t\left( \mathbf{\Theta }_r \right) 
			  $           	& The transmission (reflection) beamforming vector and diagonal matrix  		\\
			\hline
			$M,C,K$				&  	The total number of the elements of the STAR-RIS, clusters and users		\\ 
			\hline
			  $\mathbf{g}_{c,n},\mathbf{h}_{c,n}$  &   channel/combined channel vector from the STAR-RIS to user $n$ in cluster $c$ \\
			  \hline
			  $\mathbf{F}$  & channel matrix from the BS to STAR-RIS \\
			  \hline
			  $\mathcal{D} _c\left( k \right) $ & the user index of the $k$-th decoded user order \\
			\hline
			$\rho _{c,n}$, $s_{c,n}$ & power allocation coefficient and the desired signal of user $n$ in cluster $c$ \\
			\hline
			$\mathbf {w}_c$   & active beamforming vector for cluster $c$\\			
			\hline
			$\mathrm{SINR}_{j \rightarrow k}^{c}$,$\mathrm{SINR}_{k \rightarrow k}^{c}$  & the SINR for user $j $/$k $ to decode user $k $ \\
			\hline
			$\mathrm{R}_{j \rightarrow k}^{c}$,$\mathrm{R}_{k \rightarrow k}^{c}$  &  The achievable rate for user $j$/$k $ to decode user $k$ \\
			\hline
			$P_{\rm max}$, $R_{c,k}^{\rm min}$  & total transmit power budget and minmum QoS requirement \\
			\hline
		\end{tabular}
	\end{center}
\end{table*}
 \section{\textcolor[rgb]{0.00,0.00,0.00}{Physics Principle And Basic Signal Model of the STAR-RIS}}
\textcolor[rgb]{0.00,0.00,0.00}{To begin with, we provide a brief introduction of the physics principle and basic signal model of the STAR-RIS$\footnote{More details of STAR-RISs can be found in the original research work~\cite{STAR}}$.}
\subsection{\textcolor[rgb]{0.00,0.00,0.00}{ Physics Principle}}
 \vspace{-0.1cm}
    \begin{figure}[!t]
	\setlength{\abovecaptionskip}{3pt}
	\setlength{\belowcaptionskip}{-20pt}
	\centering
	\includegraphics[scale=0.12]{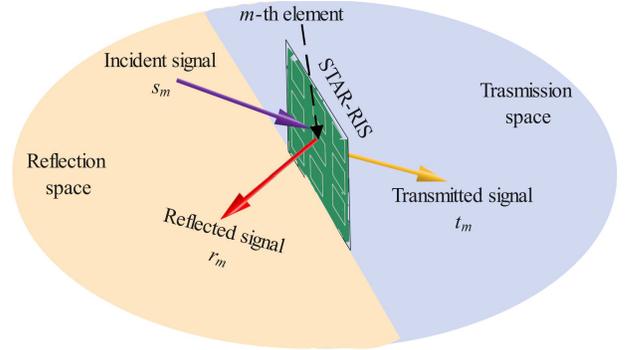}
	\caption{Signal propagation of the STAR-RIS}
	\label{STAR-RIS model}
\end{figure}
\textcolor[rgb]{0.00,0.00,0.00}{To support simultaneous transmission and reflection, the elements of STAR-RIS have to support both electric polarization currents and magnetization currents. As shown in Fig. 2(b) in~\cite{STAR}}, the physical principle of STAR-RISs is summarized as follows: firstly, the elements of STAR-RISs respond to both the electric and magnetic components of the incident field. As a result, a polarization density and a magnetization density are induced. Then, the time-varying electric polarization and magnetization currents are produced by the oscillating polarization and magnetization densities, respectively. Finally, the above time-varying currents radiate the transmitted and reflected fields back into free-space, producing phase discontinuities between the incident field and the transmitted or reflected fields.

\textcolor[rgb]{0.00,0.00,0.00}{If the electric and magnetic susceptibilities of each element are assumed to be constant, the amplitudes and phase shifts for transmission and reflection can be independently adjusted. As a result, the independent transmission and reflection beamforming to the two sides is produced, which can provide more DoFs to the communication design. However, the elements of STAR-RISs have to support both electric and magnetic currents, these elements need to be more sophisticated than those of traditional reflecting-only RISs.}
\subsection{\textcolor[rgb]{0.00,0.00,0.00}{Basic Signal Model}}
   As shown in Fig.~\ref{STAR-RIS model}, the incident signal is divided into two parts by an element of the STAR-RIS, i.e., transmitted signal in the transmission space and reflected signal in the reflection space. Assume that the STAR-RIS equips with $M$ elements and denote by $s_m$ the signal incident on the $m$-th element, where $m\in \mathcal{M} \triangleq \left\{ 1,2,\cdots ,M \right\} $. The transmitted and reflected signals by the $m$-th element can be respectively expressed as
 \begin{equation}\label{STAR-RIS signal}
   t_m=\left( \sqrt{\beta _{m}^{t}}e^{\jmath \theta _{m}^{t}} \right) s_m {~\rm and~} r_m=\left( \sqrt{\beta _{m}^{r}}e^{\jmath \theta _{m}^{r}} \right) s_m,
 \end{equation}
where $\left\{ \sqrt{\beta _{m}^{t}},\sqrt{\beta _{m}^{r}}\in \left[ 0,1 \right] \right\}$ and  $\left\{ \theta _{m}^{t},\theta _{m}^{r}\in \left[ 0,2\pi \right) \right\} 
$ are the amplitude and phase shift of the transmission and reflection coefficients of the $m$-th STAR-RIS element, respectively. 

It is noted that we can independently choose the transmission and reflection phase shifts $\left\{ \theta _{m}^{t},\theta _{m}^{r} \right\} $ from each other. However, the amplitude coefficients for transmission and reflection $\left\{ \sqrt{\beta _{m}^{t}},\sqrt{\beta _{m}^{r}} \right\} $ are coupled, since the sum of the energies of the transmitted and reflected signals has to be equal to the incident signal's energy, i.e., $\left| t_m \right|^2+\left| r_m \right|^2=\left| s_m \right|^2$, $\forall m\in \textcolor[rgb]{0.00,0.00,0.00}{ \mathcal{M} }$. To follow the above energy conservation law, the condition of $\beta _{m}^{t}+\beta _{m}^{r}=1$ should be guaranteed. It is easy to observe that each STAR-RIS element can be operated in the full transmission mode (T mode), full reflection mode (R mode)and simultaneous transmission and reflection mode (T$\&$R mode). As a result, three protocols for operating STAR-RISs are proposed in~\cite{STAR,9570143}, namely energy splitting (ES), mode switching (MS) and time switching (TS). In this paper, we only focus on the ES protocol. For expression convenience, let $\mathbf{u}_p=\left[ \sqrt{\beta _{1}^{p}}e^{j\theta _{1}^{p}},\sqrt{\beta _{2}^{p}}e^{j\theta _{2}^{p}},\cdots ,\sqrt{\beta _{M}^{p}}e^{j\theta _{M}^{p}} \right] ^H$ be the transmission $\left( p=t \right) $ or reflection $\left( p=r\right)$ beamforming vector, $\mathbf{\Theta }_p=\mathrm{diag}\left( \mathbf{u}_p^H \right) $ be the corresponding diagonal beamforming matrix. Moreover, the set of constraints to the transmission and reflection coefficients is denoted by:  
 \begin{align}\label{STAR-RISs coefficients}
 \mathcal{R} _{\beta ,\theta}=\left\{ \begin{array}{c}
 \beta _{m}^{t},\beta _{m}^{r},\\
 \theta _{m}^{t},\theta _{m}^{r}\\
 \end{array}\left| \begin{array}{c}
 \beta _{m}^{t},\beta _{m}^{r}\in \left[ 0,1 \right] ;\beta _{m}^{t}+\beta _{m}^{r}=1;\\
 \theta _{m}^{t},\theta _{m}^{r}\in \left[ 0,2\pi \right)\\
 \end{array} \right. \right\} .
 \end{align}
 \section{System Model and Problem Formulation} 
  \subsection{System Model of The Proposed STAR-RIS-NOMA system}
  \begin{figure}[!t]
  	\centering
  	\includegraphics[scale=0.1]{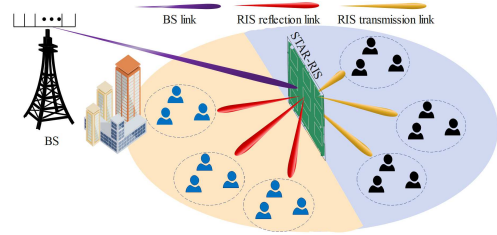}
  	\caption{STAR-RIS-NOMA system}
  	\label{system_model}
  \end{figure}
 As shown in Fig.~\ref{system_model}, we consider the downlink transmission in a STAR-RIS-NOMA communication system, where the direct BS-user links are blocked and the BS communicates with $K$ single-antenna users with the aid of a STAR-RIS. Assume that the BS is equipped with ${N_{\rm T}}$ transmit antennas, while the STAR-RIS is equipped with $M$ elements. \textcolor[rgb]{0.00,0.00,0.00}{We also assume that the \emph{K} users are grouped into \emph{C} clusters by employing user clustering techniques}$\footnote{\textcolor[rgb]{0.00,0.00,0.00}{User clustering methods \textcolor[rgb]{0.00,0.00,0.00}{have} been extensively studied for cluster-based NOMA \textcolor[rgb]{0.00,0.00,0.00}{networks}, such as \textcolor[rgb]{0.00,0.00,0.00}{K-means}~\cite{8454272}, many-to-one matching~\cite{9167258}, correlation of channels~\cite{8485639}. These methods can be applied for our considered system. The development of more sophisticated user clustering methods for STAR-RIS-NOMA systems is an interesting topic for future research. In this paper, we only focus on the power allocation and beamformnig optimization problem.}} $ . The cluster and user sets are denoted by $\textcolor[rgb]{0.00,0.00,0.00}{ \mathcal{C} } =\left\{ 1,\cdots ,C \right\}$ and $\textcolor[rgb]{0.00,0.00,0.00}{ \mathcal{K} } =\left\{ 1,\cdots ,K \right\}$, respectively. Moreover, denote by ${{\mathcal K}_c}$ the set of users in cluster $c$, where $\textcolor[rgb]{0.00,0.00,0.00}{ \mathcal{K} } =\cup _{c\in}\textcolor[rgb]{0.00,0.00,0.00}{ \mathcal{K} } _c$, $\textcolor[rgb]{0.00,0.00,0.00}{ \mathcal{K} } _c\cap \textcolor[rgb]{0.00,0.00,0.00}{ \mathcal{K} } _{\underline{c}}=\oslash \left( c,\underline{c}\in \textcolor[rgb]{0.00,0.00,0.00}{ \mathcal{C} } ,c\ne \underline{c} \right) $. Thus, the number of users in cluster $c$ can be denoted by $K_c=\left| \textcolor[rgb]{0.00,0.00,0.00}{ \mathcal{K} } _c \right|$, where $\sum_{c=1}^C{K_c}=K$. \textcolor[rgb]{0.00,0.00,0.00}{We assume that all the perfect CSI are available at the BS. The channel acquisition methods for the STAR-RIS-NOMA system are outside the scope of this work. However, our results can serve as a theoretical system performance benchmark. In addition, channel estimation methods~\cite{9130088,9366805} proposed for conventional RISs can be applied for STAR-RISs. The design of more efficient CSI estimation methods and analysis of imperfect CSI for STAR-RIS-NOMA systems are interesting but challenging new research problems.}

Denote by $\mathbf{h}_{c,n}=\mathbf{g}_{c,n}^{H}\mathbf{\Theta }_{c,n}\mathbf{F}$ the combined channel of the BS-RIS-user link for user $n$ in cluster $c$, where $\mathbf{F}\in \mathcal{C} ^{M\times N_{\mathrm{T}}}$ is the channel matrix from the BS to the RIS, $\mathbf{g}_{c,n}\in \mathcal{C} ^{M\times 1}$ is the channel vector from the RIS to user $n$, $\mathbf{\Theta }_{c,n}$ is the RIS coefficient diagonal matrix which is defined as 
\begin{align}\label{Phase_Umk}
	\mathbf{\Theta }_{c,n}= \left\{ 
	\begin{array}{rcl}
		 \mathbf{\Theta }_t,&\rm ~if~user~{\it n}~is~in~the~ transmission~space~(TS),\\
		 \mathbf{\Theta }_r,&\rm ~if~user~{\it n}~is~in~the~ reflection~space~(RS). 
	\end{array} \right.   
\end{align}

In the STAR-RIS-NOMA system, the BS broadcasts $C$ independent superposed data streams to the $K$ users with beamforming vectors $\left\{ \mathbf{w}_1,\mathbf{w}_2,\cdots ,\mathbf{w}_C \right\}$, where $\mathbf{w}_c$ is the active beamforming vector for cluster $c$. Therefore, the received signal at the $n$-th user in cluster $c$ is given by
\begin{align}\label{Received signal}
\begin{split}
	y_{c,n}= & \underset{\rm Desired~signal}{\underbrace{\mathbf{h}_{c,n}\mathbf{w}_c\sqrt{\rho _{c,n}}s_{c,n}}}+\underset{\rm Intra-cluster~interference}{\underbrace{\mathbf{h}_{c,n}\mathbf{w}_c\sum_{j\in \mathcal{K} _c,j\ne n,}{\sqrt{\rho _{c,j}}s_{c,j}}}} \\ 
	& +  \underset{\mathrm{Inter}-\mathrm{cluster}~\mathrm{interference}}{\underbrace{\mathbf{h}_{c,n}\sum_{\underline{c}\in \mathcal{C} ,\underline{c}\ne c}{\sum_{i\in \mathcal{K} _{\underline{c}}}{\mathbf{w}_{\underline{c}}\sqrt{\rho _{\underline{c},i}}s_{\underline{c},i}}}}}+\underset{\mathrm{Noise}}{\underbrace{z_{c,n}}},
	\end{split}
\end{align} 
where $\rho _{c,n}$ and $s_{c,n}$ are the power allocation coefficient and the desired signal of user $n$ in cluster $c$. The power allocation coefficients satisfy $\sum_{n\in \mathcal{K} _c}{\rho _{c,n}}=1$. In addition, $z_{c,n}$ is the complex circular i.i.d. additive Gaussian noise with $z_{c,n}\in \mathcal{C} \mathcal{N} \left( 0,\sigma ^2 \right) $, where $\sigma ^2$ is the noise power. 

Since there are more than one user in cluster $c$, according to the principle of NOMA, each user tries to remove the intra-cluster interference by using \textcolor[rgb]{0.00,0.00,0.00}{SIC$\footnote{\textcolor[rgb]{0.00,0.00,0.00}{SIC is regarded as an interference-cancellation technique~\cite{liu2018non}. SIC enables the user with the strongest signal to be detected first, who has hence the least interference-contaminated signal. Then, the strongest user reencodes and remodulates its signal. After that, the signal is then subtracted from the received signal. The same procedure is followed by the second strongest signal. Finally, the weakest user decodes its signal without suffering from any interference at all. More details of the SIC procedure in NOMA can be found in~\cite{9151196,9151208}.}}$} in a successive order. Therefore, the decoding order is an essential issue for NOMA systems. For traditional SISO NOMA systems, the optimal decoding order is determined by the channel gains. However, this decoding order method cannot be used directly in RIS-NOMA systems. This
is because the end-to-end channels can be modified by the RIS
and the decoding order can also be effected by the inter-cluster interference. Therefore, the optimal decoding order for cluster $c$ can be any one of the $K_c!$ different orders. Without loss of generality, denote by $\mathcal{D} _c\left( k \right) $ the user index that corresponds to the $k$-th decoded user order in cluster $c$. After applying the SIC decoding procedure\textcolor[rgb]{0.00,0.00,0.00}{~\cite{8454272}}, the received signal-to-interference-plus-noise ratio (SINR) at user $\mathcal{D} _c\left( k \right) $ can be expressed as  
\begin{align}\label{SINRkk}
\begin{split}
	& \mathrm{SINR}_{\mathcal{D} _c\left( k \right) \rightarrow \mathcal{D} _c\left( k \right)}^{c}= \\
	&\frac{\left| \mathbf{h}_{c,\mathcal{D} _c\left( k \right)}\mathbf{w}_c \right|^2\rho _{c,\mathcal{D} _c\left( k \right)}}{\left| \mathbf{h}_{c,\mathcal{D} _c\left( k \right)}\mathbf{w}_c \right|^2\sum_{n>k}{\rho _{c,\mathcal{D} _c\left( n \right)}}+\sum_{\underline{c}\ne c}{\left| \mathbf{h}_{c,\mathcal{D} _c\left( k \right)}\mathbf{w}_{\underline{c}} \right|^2}+\sigma ^2},
\end{split}
\end{align} 
where $n\in \mathcal{K} _c$ and $\underline{c}\in \mathcal{C}$. The corresponding achievable data rate is $R_{\mathcal{D} _c\left( k \right) \rightarrow \mathcal{D} _c\left( k \right)}^{c}=\log _2\left( 1+\mathrm{SINR}_{\mathcal{D} _c\left( k \right) \rightarrow \mathcal{D} _c\left( k \right)}^{c} \right) 
$. 

For any two users $\mathcal{D} _c\left( j \right) $ and $\mathcal{D} _c\left( k \right) $ with decoding order $j>k$, the SINR for user $\mathcal{D} _c\left( j \right) $ to decode user $\mathcal{D} _c\left( k \right) $ is given by
\begin{align}\label{SINRjk}
\begin{split}
&	\mathrm{SINR}_{\mathcal{D} _c\left( j \right) \rightarrow \mathcal{D} _c\left( k \right)}^{c}= \\
&\frac{\left| \mathbf{h}_{c,\mathcal{D} _c\left( j \right)}\mathbf{w}_c \right|^2\rho _{c,\mathcal{D} _c\left( k \right)}}{\left| \mathbf{h}_{c,\mathcal{D} _c\left( j \right)}\mathbf{w}_c \right|^2\sum_{n>k}{\rho _{c,\mathcal{D} _c\left( n \right)}}+\sum_{\underline{c}\ne c}{\left| \mathbf{h}_{c,\mathcal{D} _c\left( j \right)}\mathbf{w}_{\underline{c}} \right|^2}+\sigma ^2},
\end{split}
\end{align}
and the corresponding achievable rate is $R_{\mathcal{D} _c\left( j \right) \rightarrow \mathcal{D} _c\left( k \right)}^{c}=\log _2\left( 1+\mathrm{SINR}_{\mathcal{D} _c\left( j \right) \rightarrow \mathcal{D} _c\left( k \right)}^{c} \right) $. 

To guarantee the SIC performed successfully, the condition $R_{\mathcal{D} _c\left( j \right) \rightarrow \mathcal{D} _c\left( k \right)}^{c}\geqslant R_{\mathcal{D} _c\left( k \right) \rightarrow \mathcal{D} _c\left( k \right)}^{c}
$ with $j>k$ should be satisfied. For example, if there are three users in cluster $c$. Then, the following SIC decoding rate conditions at user  $\mathcal{D} _c\left( 2 \right) $ and  $\mathcal{D} _c\left( 3 \right) $ should be kept:
\begin{equation}\label{SIC condition}
\left\{ \begin{array}{c}
R_{\mathcal{D} _c\left( 2 \right) \rightarrow \mathcal{D} _c\left( 1 \right)}^{c}\geqslant R_{\mathcal{D} _c\left( 1 \right) \rightarrow \mathcal{D} _c\left( 1 \right)}^{c},\\
R_{\mathcal{D} _c\left( 3 \right) \rightarrow \mathcal{D} _c\left( 2 \right)}^{c}\geqslant R_{\mathcal{D} _c\left( 2 \right) \rightarrow \mathcal{D} _c\left( 2 \right)}^{c},\\
R_{\mathcal{D} _c\left( 3 \right) \rightarrow \mathcal{D} _c\left( 1 \right)}^{c}\geqslant R_{\mathcal{D} _c\left( 1 \right) \rightarrow \mathcal{D} _c\left( 1 \right)}^{c}.\\
\end{array} \right. 
\end{equation} 

As a result, there will be $\frac{K_c\left( K_c-1 \right)}{2}$ SIC decoding rate conditions for cluster $c$ with $K_c$ users. It is worth noting that the SIC decoding rate conditions depend not only on the active beamforming vectors and power allocation coefficients, but also on the transmission and reflection beamforming vectors.

Finally, the overall achievable sum rate of the proposed STAR-RIS-NOMA system can be written as
\begin{equation}\label{sum rate}
	R_{\mathrm{sum}}=\sum_{c\in \mathcal{C}}{\sum_{k\in \mathcal{K} _c}{R_{\mathcal{D} _c\left( k \right) \rightarrow \mathcal{D} _c\left( k \right)}^{c}}}.
\end{equation}
\vspace{-1cm}
\subsection{Problem Formulation}
To improve the overall data rate, we formulate a joint decoding order, power allocation coefficients, active beamforming, transmission and reflection beamforming optimization problem to maximize the achievable sum rate of the $K$ users. The problem is formulated as
\begin{subequations}\label{OP1}
	\begin{align}
	&\underset{\mathcal{D}_c ,\rho _{c,\mathcal{D} _c\left( k \right)}
		,\mathbf{w}_c,\mathbf{u}_p}{\max}\sum_{c\in \mathcal{C}}{\sum_{k\in \mathcal{K} _c}{R_{\mathcal{D} _c\left( k \right) \rightarrow \mathcal{D} _c\left( k \right)}^{c}}}, \\
	&s.t.~R_{\mathcal{D} _c\left( k \right) \rightarrow \mathcal{D} _c\left( k \right)}^{c}\geqslant R_{c,\mathcal{D} _c\left( k \right)}^{\min},  \label{OP1:b} \\
	&   \ \ \ \ \ R_{\mathcal{D} _c\left( j \right) \rightarrow \mathcal{D} _c\left( k \right)}^{c}\geqslant R_{\mathcal{D} _c\left( k \right) \rightarrow \mathcal{D} _c\left( k \right)}^{c},j\geqslant k,  \label{OP1:c} \\
	&   \ \ \ \ \ \sum_{c\in \mathcal{C}}{\left\| \mathbf{w}_c \right\| _{2}^{2}}\leqslant P_{\max}, \label{OP1:d} \\
	&   \ \ \ \ \ \sum_{k\in \mathcal{K} _c}{\rho _{c,\mathcal{D} _c\left( k \right)}}=1,  \label{OP1:e} \\
	&   \ \ \ \ \ \beta _{m}^{p},\theta _{m}^{p}\in \mathcal{R} _{\beta ,\theta}, \label{OP1:f}  \\
	&   \ \ \ \ \ \mathcal{D} _c\in \mathcal{D},   c\in \mathcal{C}, \label{OP1:g}
	\end{align}
\end{subequations}
where $ k,~j\in \mathcal{K}_c,~c\in \mathcal{C},~m\in \mathcal{M},~p\in \left\{ t,r \right\}$. Constraint~\eqref{OP1:b} ensures the minimum QoS requirement of each user, constraint~\eqref{OP1:c} guarantees the success of the SIC decoding, constraint~\eqref{OP1:d} indicates that the total transmit power budget is $P_{\rm max}$, constraint~\eqref{OP1:e} represents the power allocation coefficient constraint in each cluster, constraint~\eqref{OP1:f} is for the amplitude and phase shift coefficients of the STAR-RIS. In constraint~\eqref{OP1:g}, $\mathcal{D}$ denotes the combination set of all possible decoding orders.

Compared to the traditional RIS with only reflection coefficients, the new introduced transmission and reflection coefficients by employing the STAR-RIS are highly coupled. Thus, the formulated problem~\eqref{OP1} is more challenging to solve than that for traditional RIS. It is noted that the optimization problem for traditional RISs is a special case of the optimization problem for STAR-RISs, where the transmission function is turned off and only the reflection function can work. In the following sections, we will develop a new algorithm to decouple the optimization variables.
\section{Solution of the Problem}
To solve problem~\eqref{OP1}, a new iterative algorithm is  proposed, which includes two layers, i.e., inner layer and outer layer. The outer-layer iteration is designed to determine the decoding order. With the obtained decoding order, the joint optimization problem over power allocation coefficients, active beamforming, transmission and reflection beamforming are solved by the inner-layer iteration.
\subsection{Equivalent-Combined Channel Gain based Decoding Order}
\textcolor[rgb]{0.00,0.00,0.00}{Decoding order is an essential problem for the considered STAR-RIS-NOMA system and should be determined before solving the optimization problem.} Therefore, we first present a new scheme to determine the decoding order by introducing the following lemma.   
\begin{lemma}\label{decoding order}
	For cluster $c$ with $K_c$ users, under given active beamforming vectors $\left\{ \mathbf{w}_c \right\}$ , transmission beamforming vector $\mathbf{u}_t $ and reflection beamforming vector $\mathbf{u}_r$, the optimal decoding order is defined as~\cite{8454272,Wang2019StackelbergGF}
	\begin{align}\label{optimal decoding oder}
	\Gamma _{c,\mathcal{D} _c\left( 1 \right)}\leqslant \Gamma _{c,\mathcal{D} _c\left( 2 \right)}\leqslant \cdots \leqslant \Gamma _{c,\mathcal{D} _c\left( K_c \right)},
	\end{align}
	where $\Gamma_{c,\mathcal{D} _c\left( k \right)}$ is the equivalent-combined channel gain, which can be expressed as follows
	\begin{align}\label{equivalent channel gain}
		\Gamma _{c,\mathcal{D} _c\left( k \right)}=\frac{\left| \mathbf{g}_{c,\mathcal{D} _c\left( k \right)}^{H}\mathbf{\Theta }_{c,\mathcal{D} _c\left( k \right)}\mathbf{Fw}_c \right|^2}{\sum_{\underline{c}\in \mathcal{C} ,\underline{c}\ne c}{\left| \mathbf{g}_{c,\mathcal{D} _c\left( k \right)}^{H}\mathbf{\Theta }_{c,\mathcal{D} _c\left( k \right)}\mathbf{Fw}_{\underline{c}} \right|^2}+\sigma ^2},
	\end{align}   
\end{lemma}

\textbf{Lemma~\ref{decoding order}} indicates that the decoding order for
each cluster of the STAR-RIS assisted NOMA system is a function of the active beamforming vectors $\left\{ \mathbf{w}_c \right\} $, transmission and reflection beamforming vectors $\left\{ \mathbf{u }_p \right\}$. The power allocation coefficients $\left\{ \rho _{c,\mathcal{D} _c\left( k \right)} \right\}$ has no impact on the decoding order.
\begin{proposition} \label{proposition of lemma 1}
	For any two users $k$ and $j$ belong to cluster $c$, 
	if the decoding order of the two users satisfies 
	\begin{equation}
	  \mathcal{D} _{c}^{-1}\left( j \right) >\mathcal{D} _{c}^{-1}\left( k \right),
	\end{equation}
	 where $\mathcal{D} _{c}^{-1}\left( \cdot \right)$ is the inverse of mapping function $\mathcal{D} _c\left( \cdot \right) $. 
	 
	 Then, under the optimal decoding order, the following SIC condition is guaranteed:
	\begin{align}\label{SICi-j}
		R_{j\rightarrow k}^{c}\geqslant R_{k\rightarrow k}^{c}.
	\end{align} 
	Proof: See Appendix A.
\end{proposition}

\textbf{Proposition~\ref{proposition of lemma 1}} indicates that the SIC constraints in~\eqref{OP1:c} can be removed under the optimal decoding order. However, this operation will not affect the optimality of problem~\eqref{OP1}. Based on this observation, we propose a suboptimal algorithm to solve problem~\eqref{OP1} by iteratively updating the optimal decoding order and solving problem~\eqref{OP1} with no SIC constraints.

Without loss of generality, let $\mathcal{D} _c\left( k \right) =k$. Then, problem~\eqref{OP1} under the given decoding order can be rewritten as:
\begin{subequations}\label{OP2}
	\begin{align}
	&\underset{\rho _{c,k},\mathbf{w}_c,\mathbf{u}_p}{\max}\sum_{c\in \mathcal{C}}{\sum_{k\in \mathcal{K} _c}{R_{k\rightarrow k}^{c}}}
	, \\
	&s.t.~R_{k\rightarrow k}^{c}\geqslant R_{c,k}^{\min}, \forall k\in \mathcal{K} _c,\forall c\in \mathcal{C} 
	, \label{OP2:b} \\
	&   \ \ \ \ \sum_{k\in \mathcal{K} _c}{\rho _{c,k}}=1, \forall c\in \mathcal{C} , \label{OP2:c}\\
	&   \ \ \ \ \ \eqref{OP1:d},\eqref{OP1:f}. \label{OP2:d}  
	\end{align}
	
	In the following, we solve the reduced problem~\eqref{OP2} instead of the original problem~\eqref{OP1}.
\end{subequations}
 \vspace{-0.5cm}
\subsection{Power Allocation Coefficients Optimization }
We assume that the active beamforming vectors $\left\{ \mathbf{w}_c \right\} $, transmission and reflection beamforming vectors $\left\{ \mathbf{u}_p \right\} $ are known. Then, the optimal decoding order can be obtained according to \textbf{Lemma~\ref{decoding order}}. Since the inter-cluster interference has no relationship with the power allocation coefficients $\left\{ \rho _{c,k} \right\}$, the optimization problem in~\eqref{OP2} for the power allocation coefficients can be decomposed into $C$ decoupled subproblems. Without loss of generality, for any cluster $c$, the power allocation coefficients optimization problem is simplified as:
\begin{subequations}\label{power allocation}
	\begin{align}
	&\underset{ \rho _{c,k}}{\max}\sum_{k\in \mathcal{K} _c}{R_{k\rightarrow k}^{c}}, \\
	&s.t.~\eqref{OP2:b},\eqref{OP2:c}. \label{PA:d}  
	\end{align}
\end{subequations}
\begin{lemma}\label{feasible}
	Given the active beamforming vectors $\left\{ \mathbf{w}_c \right\} $, transmission and reflection beamforming vectors $\left\{ \mathbf{u}_p \right\} $ and decoding order $\mathcal{D} _c$, if the following inequality holds:
	\begin{equation}\label{feasible rho}
		\sum_{k=1}^{K_c}{\frac{r_{c,k}^{\min}}{\Gamma _{c,k}}\prod_{i=1}^{k-1}{\left( r_{c,i}^{\min}+1 \right)}}\leqslant 1,
	\end{equation}
where \textcolor[rgb]{0.00,0.00,0.00}{$r_{c,k}^{\min}=2^{R_{c,k}^{\min}}-1$.}

Then, problem~\eqref{power allocation} is feasible.
	
Proof: See Appendix B.	
\end{lemma}
\begin{theorem}\label{Theorem optimal power}
 If problem~\eqref{power allocation} is feasible, its optimal objective value is given by
\begin{equation}\label{max sum rate}
	R_{\mathrm{sum}}=\sum_{c=1}^C{\sum_{k=2}^{K_c-1}{R_{c,k}^{\min}}}+\sum_{c=1}^C{\log _2\left( 1+\rho _{c,K_c}^{*}\Gamma _{c,K_c} \right)},
\end{equation}
 and the optimal power allocation coefficients can be expressed as
 
 \begin{equation} \label{optimal power}
 	 \setlength{\abovedisplayskip}{0pt}
 	\setlength{\belowdisplayskip}{0pt}
 \begin{cases}
 	\rho _{c,1}^{*}=\frac{r_{c,1}^{\min}}{1+r_{c,1}^{\min}}\left( 1+\frac{1}{\Gamma _{c,1}} \right),\\
 \rho _{c,2}^{*}=\frac{r_{c,2}^{\min}}{1+r_{c,2}^{\min}}\left( 1-\rho _{c,1}^{*}+\frac{1}{\Gamma _{c,2}} \right),\\
 \vdots\\
 \rho _{c,K_c-1}^{*}=\frac{r_{c,K_c-1}^{\min}}{1+r_{c,K_c-1}^{\min}}\left( 1-\sum_{i=1}^{K_c-2}{\rho _{c,i}^{*}}+\frac{1}{\Gamma _{c,K_c-1}} \right),\\
 \rho _{c,K_c}^{*}=1-\sum_{i=1}^{K_c-1}{\rho _{c,i}^{*}}.
 \end{cases}
 \end{equation}
Proof: See Appendix C.
 \vspace{-0.5cm}
\end{theorem}
\subsection{Active Beamforming Optimization}
In this subsection, we focus on the active beamforming optimization problem in~\eqref{OP2} with given power allocation coefficients $\left\{ \rho _{c,k} \right\} $, transmission and reflection beamforming vectors $\left\{  \textbf{u}_p \right\}$. Before solving this problem, we introduce a slack variable set $\left\{ A_{c,k},B_{c,k}|k\in \mathcal{K} _c,c\in \mathcal{C} \right\} 
$, where $ A_{c,k}$ and $B_{c,k}$ are defined as 
\begin{equation}\label{Ack}
	 \setlength{\abovedisplayskip}{0pt}
	\setlength{\belowdisplayskip}{0pt}
\frac{1}{A_{c,k}}=\left| \mathbf{h}_{c,k}\mathbf{w}_c \right|^2\rho _{c,k},
\end{equation}
\begin{equation}\label{Bck}
	 \setlength{\abovedisplayskip}{0pt}
	\setlength{\belowdisplayskip}{0pt}
B_{c,k} = \left| \mathbf{h}_{c,k}\mathbf{w}_c \right|^2\sum_{n\in \mathcal{K} _c,n>k}{\rho _{c,n}}+\sum_{\underline{c}\in \mathcal{C} ,\underline{c}\ne c}{\left| \mathbf{h}_{{c},k}\mathbf{w}_{\underline{c}} \right|^2}+\sigma ^2.
\end{equation}

Substituting~\eqref{Ack} and~\eqref{Bck} into~\eqref{SINRkk}, the achievable data rate can be rewritten as
\begin{equation}\label{rewritten Rkk1}
	 \setlength{\abovedisplayskip}{0pt}
	\setlength{\belowdisplayskip}{0pt}
	R_{k\rightarrow k}^{c}=\log _2\left( 1+\frac{1}{A_{c,k}B_{c,k}} \right), 
\end{equation}

Combining with~\eqref{Ack},~\eqref{Bck} and~\eqref{rewritten Rkk1}, the active beamforming optimization problem in~\eqref{OP2} can be expressed as

\begin{subequations}\label{OP3}
 \setlength{\abovedisplayskip}{0pt}
  \setlength{\belowdisplayskip}{0pt}
	\begin{align}
	&\underset{\mathbf{w}_c,{A}_{c,k},{B}_{c,k},\textcolor[rgb]{0.00,0.00,0.00}{R_{k\rightarrow k}^{c}}}{\max}\sum_{c\in \mathcal{C}}{\sum_{k\in \mathcal{K} _c}{R_{k\rightarrow k}^{c}}}
	, \\
	&s.t.~\log _2\left( 1+\frac{1}{{A}_{c,k}{B}_{c,k}} \right) \geqslant R_{k\rightarrow k}^{c}, \label{OP3:b} \\
	&   \ \ \ \ \frac{1}{{A}_{c,k}}\leqslant \left| \mathbf{h}_{c,k}\mathbf{w}_c \right|^2\rho _{c,k},\label{OP3:c}\\
	&   \ \ \ \ {B} _{c,k}\geqslant \left| \mathbf{h}_{c,k}\mathbf{w}_c \right|^2\sum_{n>k}{\rho _{c,n}}+\sum_{\underline{c}\ne c}{\left| \mathbf{h}_{{c},k}\mathbf{w}_{\underline{c}} \right|^2}+\sigma ^2,\label{OP3:d}\\	
	&   \ \ \ \ \ \eqref{OP1:d},~\eqref{OP2:b}, \label{OP3:e}  
	\end{align}
\end{subequations}
 where $c, \underline{c}\in \mathcal{C}$ and $k, n\in \mathcal{K}_c$.
 
We further define $\mathbf{H}_{c,k}=\mathbf{h}_{c,k}^{H}\mathbf{h}_{c,k}
$ and $\mathbf{W}_c=\mathbf{w}_c\mathbf{w}_{c}^{H}$, where $\mathbf{W}_c\succeq 0$ and $ \rm rank\left( \mathbf{W}_c \right) =1$. Then, we have:$\left| \mathbf{h}_{c,k}\mathbf{w}_c \right|^2=\mathrm{Tr}\left( \mathbf{W}_c\mathbf{H}_{c,k} \right) 
$. Finally, problem~\eqref{OP3} can be reformulated as:
\begin{subequations}\label{ABO}
 \setlength{\abovedisplayskip}{0pt}
	\setlength{\belowdisplayskip}{0pt}
	\begin{align}
		&\underset{\mathbf{W}_c,{A}_{c,k}, {B}_{c,k},\textcolor[rgb]{0.00,0.00,0.00}{R_{k\rightarrow k}^{c}}}{\max} \sum_{c\in \textcolor[rgb]{0.00,0.00,0.00}{ \mathcal{C} }}{\sum_{k\in \textcolor[rgb]{0.00,0.00,0.00}{ \mathcal{K} } _c}{R_{k\rightarrow k}^{c}}} 	, \\
		&s.t.~\frac{1}{{A}_{c,k}}\leqslant \mathrm{Tr}\left( \mathbf{W}_m\mathbf{H}_{c,k} \right) \rho _{c,k}, \label{ABO:b} \\
		&   \ \ \ \ \begin{array}{l} B_{c,k}\geqslant \mathrm{Tr}\left( \mathbf{W}_c\mathbf{H}_{c,k} \right) \sum_{n\in \textcolor[rgb]{0.00,0.00,0.00}{ \mathcal{K} } _c,n>k}{\rho _{c,n}} \\ \ \ \ \ \ \ \ \ \ +\sum_{\underline{c}\in \textcolor[rgb]{0.00,0.00,0.00}{ \mathcal{C} } ,\underline{c}\ne c}{\mathrm{Tr}\left( \mathbf{W}_{\underline{c}}\mathbf{H}_{c,k} \right)}+\sigma ^2
		,\end{array} \label{ABO:c}\\
		&   \ \ \ \ \sum_{c\in \textcolor[rgb]{0.00,0.00,0.00}{ \mathcal{C} }}{\mathrm{Tr}\left( \mathbf{W}_c \right)}\leqslant P_{\max},\label{ABO:d}\\	
		&   \ \ \ \ \ \mathrm{rank}\left( \mathbf{W}_c \right) =1,\label{ABO:e}\\	
		&   \ \ \ \ \  	\mathbf{W}_c\succcurlyeq 0,\label{ABO:f}\\	
		&   \ \ \ \ \eqref{OP2:b},~\eqref{OP3:b}, \label{ABO:g}  
	\end{align}
\end{subequations}
where $c\in \textcolor[rgb]{0.00,0.00,0.00}{ \mathcal{C} }$ and $k\in \textcolor[rgb]{0.00,0.00,0.00}{ \mathcal{K} }_c$.

\textcolor[rgb]{0.00,0.00,0.00}{It is noted that $\log _2\left( 1+\frac{1}{xy} \right) $ is a joint convex function with respect to $x$ and $y$~\cite{9139273}. Therefore, the left-hand term of constraint~\eqref{OP3:b}, i.e.,  $\log _2\left( 1+\frac{1}{{A}_{c,k}{B}_{c,k}} \right)$, is also joint convex function over ${A}_{c,k}$ and ${B}_{c,k}$, which makes constraint~\eqref{OP3:b} non-convex. As a result, problem~\eqref{ABO} is a non-convex optimization problem due to the non-convex constraints~\eqref{OP3:b} and~\eqref{ABO:e}. To begin with, we approximate $\log _2\left( 1+\frac{1}{{A}_{c,k}{B}_{c,k}} \right)$ by applying the first-order Taylor expansion. Thus, the lower bound can be expressed as in Equation~\eqref{Taylor} at the top of next page, where the points $A_{c,k}^{\left( \tau_1 \right)}
	$ and $B_{c,k}^{\left( \tau_1 \right)}
	$ are the values of $A_{c,k}
	$ and $B_{c,k}$ in the $\tau_1$-th iteration, respectively.  $\widetilde{R}_{k\rightarrow k}^{\mathrm{c}}$ is a linear function over ${A}_{c,k}$ and ${B}_{c,k}$ and is lower bound of function $\log _2\left( 1+\frac{1}{{A}_{c,k}{B}_{c,k}} \right)$.}
  \begin{figure*}[!t]
 	\normalsize
 \begin{equation} \label{Taylor} 
 \begin{split}
	\log _2\left( 1+\frac{1}{A_{c,k}B_{c,k}} \right) & \geqslant \log _2\left( 1+\frac{1}{A_{c,k}^{\left( \tau_1 \right)}B_{c,k}^{\left( \tau_1 \right)}} \right) -\frac{\log _2\mathrm{e}\left( A_{c,k}-A_{c,k}^{\left( \tau_1 \right)} \right)}{A_{c,k}^{\left(\tau_1 \right)}\left( 1+B_{c,k}^{\left( \tau_1 \right)}B_{c,k}^{\left( \tau_1 \right)} \right)}-\frac{\log _2\mathrm{e}\left( B_{c,k}-B_{c,k}^{\left(\tau_1 \right)} \right)}{A_{c,k}^{\left( \tau_1 \right)}\left( 1+A_{c,k}^{\left( \tau_1 \right)}B_{c,k}^{\left( \tau_1 \right)} \right)}\\
	&=\widetilde{R}_{k\rightarrow k}^{\mathrm{c}},
 \end{split}
 \end{equation}  
 	\hrulefill \vspace*{0pt}
\end{figure*} 
 
 \textcolor[rgb]{0.00,0.00,0.00}{As a result, constraint~\eqref{OP3:b} can be approximated by the convex constraint $\widetilde{R}_{k\rightarrow k}^{\mathrm{c}}\geqslant R_{k\rightarrow k}^{c}$ in the $\tau_1$-th iteration. According to SCA algorithm, solving problem~\eqref{ABO} is equivalent to iteratively solving the following problem
 	\begin{subequations}\label{ABO_1}
 		\setlength{\belowdisplayskip}{2pt}
 		\begin{align}
 			&\underset{\mathbf{W}_c,{A}_{c,k}, {B}_{c,k},\textcolor[rgb]{0.00,0.00,0.00}{R_{k\rightarrow k}^{c}}}{\max} \sum_{c\in \textcolor[rgb]{0.00,0.00,0.00}{ \mathcal{C} }}{\sum_{k\in \textcolor[rgb]{0.00,0.00,0.00}{ \mathcal{K} } _c}{R_{k\rightarrow k}^{c}}} 	, \\
 			&s.t.~\widetilde{R}_{k\rightarrow k}^{\mathrm{c}}\geqslant R_{k\rightarrow k}^{c}, c\in \textcolor[rgb]{0.00,0.00,0.00}{ \mathcal{C} }, k\in \textcolor[rgb]{0.00,0.00,0.00}{ \mathcal{K} }_c,\label{ABO_1:b} \\
 			&   \ \ \ \ ~\eqref{OP2:b}, \eqref{ABO:b},~\eqref{ABO:c},~\eqref{ABO:d},~\eqref{ABO:e},~\eqref{ABO:f}. \label{ABO_1:c} 
 		\end{align}
 	\end{subequations} 
 }
 
 Now, the non-convex rank-one constraint~\eqref{ABO:e} is the remaining obstacle to solve problem~\eqref{ABO_1}. To solve this problem, we have the following theorem. 
 \begin{theorem}\label{rank 1}
 	The optimal $\left\{ \mathbf{W}_{c}^{*} \right\} $ to problem~\eqref{ABO_1} without the rank-one constraint~\eqref{ABO:e} always satisfy $\mathrm{rank}\left( \mathbf{W}_c \right) =1$.
 \end{theorem}	
 \textit{Proof}: See Appendix D.
 	
 Theorem~\ref{rank 1} represents the fact that we can obtain the
 optimal $\left\{ \mathbf{W}_{c}^{*} \right\} $ of problem~\eqref{ABO} by solving problem~\eqref{ABO_1} without the rank-one constraint. Problem~\eqref{ABO_1} is a standard convex SDP, which can be solved efficiently by numerical solves such as the SDP solver in CVX tool~\cite{cvx}. Due to the replacement of the lower bound~\eqref{Taylor}, problem~\eqref{ABO_1} is a lower bound approximation of the active beamforming problem~\eqref{OP3}. {Algorithm~\ref{Active algorithm}} summarizes the proposed SCA based algorithm to solve problem~\eqref{OP3}. It is noted that  {Algorithm~\ref{Active algorithm}} is guaranteed to converge to a locally optimal solution of~\eqref{OP3}~\cite{SCA_converge}.
  \begin{algorithm} 	
 	\caption{Successive Convex Approximation (SCA) Based Algorithm for obtaining $\left\{ \mathbf{W}_{c}^{*} \right\} $}
 	\label{Active algorithm}
 	\begin{algorithmic}[1]
 		\STATE  Initialize feasible points $\left\{ A_{c,k}^{\left( 0 \right)} \right\}$,  $\left\{ B_{c,k}^{\left( 0 \right)} \right\}$ and set the iteration index ${\tau_1} = 0$.
 		\REPEAT
 		\STATE  update $\left\{ A_{c,k}^{\left( \tau_1+1 \right)} \right\}$,  $\left\{ B_{c,k}^{\left( \tau_1+1 \right)} \right\}$ and $\left\{ \textbf{W}^{\left( \tau_1+1 \right)} \right\}$ by solving problem~\eqref{ABO_1};
 		\STATE  ${\tau_1} = {\tau_1} + 1$;
 		\UNTIL {the objective value of problem~\eqref{ABO_1} converge.}
 		\STATE   \textbf{Output}: $\mathbf{W}_{c}^{*}$ 
 	\end{algorithmic}
 \end{algorithm}
 \subsection{Transmission and Reflection Beamforming Optimization}
 When the active beamforming vectors $\left\lbrace \textbf{w}_c\right\rbrace $ are given, we denote $\overline{\mathbf{h}}_{c,k,\underline{c}}=\mathrm{diag}\left( \mathbf{g}_{c,k}^{H} \right) \mathbf{Fw}_{\underline{c}}
 $ and $\mathbf{U}_{p}=\mathbf{u}_{p}\mathbf{u}_{p}^{H}$, where $\mathbf{U}_p\succeq 0$, $ \rm rank\left( \mathbf{U}_p \right) =1$ and $\left[ \mathbf{U}_p \right] _{m,m}=\beta _{m}^{p}, p\in \left\{ t,r \right\}$. Hence, we have:
 \begin{equation} \label{Trace passive} 	
	\left| \mathbf{g}_{c,k}^{H}\mathbf{\Theta }_{c,k}\mathbf{Fw}_{\underline{c}} \right|^2=\left| \mathbf{v}_{c,k}^{H}\overline{\mathbf{h}}_{c,k,\underline{c}} \right|^2=\mathrm{Tr}\left( \mathbf{V}_{c,k}\overline{\mathbf{H}}_{c,k,\underline{c}} \right), 
 \end{equation}  	
 where $\overline{\mathbf{H}}_{c,k,\underline{c}}=\overline{\mathbf{h}}_{c,k,\underline{c}}\overline{\mathbf{h}}_{c,k,\underline{c}}^{H} $ and the matrix(vector) $\mathbf{V}_{c,k}\left( \mathbf{v}_{c,k} \right)$ is defined as:
  \begin{equation} \label{Matrix define} 
	\mathbf{V}_{c,k}\left( \mathbf{v}_{c,k} \right) = \left\{ 
	\begin{array}{rcl}
		\mathbf{U}_t\left( \mathbf{u}_t \right),&  &\rm ~if~user~{\it k}~is~in~TS, \\
		\mathbf{U}_r\left( \mathbf{u}_r \right),&  &\rm ~if~user~{\it k}~is~in~RS. 
	\end{array} \right. 
  \end{equation}  
 
Thus, the transmission and reflection beamforming optimization problem in~\eqref{OP2} with fixed active beamforming vectors and power allocation coefficients can be formulated as 	
 	\begin{subequations}\label{PB}
 		\begin{align}
 		&\underset{\mathbf{U}_p,A_{c,k},B_{c,k},\textcolor[rgb]{0.00,0.00,0.00}{R_{k\rightarrow k}^{c}}}{\max}\sum_{c\in \mathcal{C}}{\sum_{k\in \mathcal{K} _c}{R_{k\rightarrow k}^{c}}}, \\
 		&s.t.~\frac{1}{A_{c,k}}\leqslant \mathrm{Tr}\left( \mathbf{V}_{c,k}\overline{\mathbf{H}}_{c,k,c} \right) \rho _{c,k}, \label{PB:b} \\
 		&   \ \ \ \ \  \begin{array}{l} B_{c,k}\geqslant \mathrm{Tr}\left( \mathbf{V}_{c,k}\overline{\mathbf{H}}_{c,k,c} \right) \sum_{ n>k}{\rho _{c,n}} \\ ~~~~~~~~~+\sum_{ \underline{c}\ne c}{\mathrm{Tr}\left( \mathbf{V}_{c,k}\overline{\mathbf{H}}_{c,k,\underline{c}} \right)}  +\sigma^{2},\end{array} \label{PB:c}\\
 		&   \ \ \ \ \ \beta _{m}^{t}+\beta _{m}^{r}=1,\label{PB:d}\\
 		&   \ \ \ \ \  \left[ \mathbf{U}_p \right] _{m,m}=\beta _{m}^{p},  \label{PB:e}  \\
 		&   \ \ \ \ \  \mathbf{U}_p\succcurlyeq 0, \label{PB:f} \\ 		
		&   \ \ \ \ \  \mathrm{rank}\left( \mathbf{U}_p \right) =1,  \label{PB:g} \\
		&   \ \ \ \ \  \eqref{OP2:b},\eqref{OP3:b}, 	
 		\end{align}
 	\end{subequations}
where $c,~\underline{c}\in \mathcal{C}$, $k,~n\in \mathcal{K}_c$,  $m\in \mathcal{M}$ and $p\in \left\{ t,r \right\}$.
 	
 According to~\cite{9139273,9423667}, the non-convex rank-one constraint~\eqref{PB:g} can be replaced by the following relaxed convex constraint:
 \begin{equation} \label{rank one relax} 
 	\varepsilon _{\max}\left( \mathbf{U}_p \right) \geqslant \epsilon ^{\left( \tau _2 \right)}\mathrm{Tr}\left( \mathbf{U}_p \right) ,
\end{equation}  	
 where $\varepsilon _{\max}\left( \mathbf{U}_p \right)$	denotes the maximum eigenvalue of matrix $\mathbf{U}_p$, $\epsilon ^{\left( \tau _2 \right)}$ is a relaxation parameter in the $\tau_2$-th iteration, which controls $\varepsilon _{\max}\left( \mathbf{U}_p \right)$ to trace ratio of $\mathbf{U}_p$. Specifically, $\epsilon ^{\left( \tau _2 \right)}=0$ indicates that the rank-one constraint is dropped; $\epsilon ^{\left( \tau _2 \right)}=1$ is equivalent to the rank-one constraint. Therefore, we can increase $\epsilon ^{\left( \tau _2 \right)}$ from 0 to 1 sequentially via iterations to gradually approach a rank-one solution. It is noted that $\varepsilon _{\max}\left( \mathbf{U}_p \right)$ is not differentiable, i.e., non-smooth. The following approximation expression can be used to approximate $\varepsilon _{\max}\left( \mathbf{U}_p \right)$:
  \begin{equation} \label{rank one relax} 
	\varepsilon _{\max}\left( \mathbf{U}_p \right) =\mathbf{e}_{\max}^{H}\left( \mathbf{U}_{p}^{\left( \tau _2 \right)} \right) \mathbf{U}_p\mathbf{e}_{\max}\left( \mathbf{U}_{p}^{\left( \tau _2 \right)} \right), 
 \end{equation}
 where $\mathbf{e}_{\max}\left( \mathbf{U}_{p}^{\left( \tau _2 \right)} \right) $ is the
 eigenvector corresponding to the maximum eigenvalue of $\mathbf{U}_{p}^{\left( \tau _2 \right)}$.
 
 Thus, solving problem~\eqref{PB} is transformed to solve the following relaxed problem
	\begin{subequations}\label{PB_relaxed}
	\begin{align}
	&\underset{\mathbf{U}_p,A_{c,k},B_{c,k},\textcolor[rgb]{0.00,0.00,0.00}{R_{k\rightarrow k}^{c}}}{\max}\sum_{c\in \mathcal{C}}{\sum_{k\in \mathcal{K} _c}{R_{k\rightarrow k}^{c}}}, \\
	&s.t.~ \mathbf{e}_{\max}^{H}\left( \mathbf{U}_{p}^{\left( \tau _2 \right)} \right) \mathbf{U}_p\mathbf{e}_{\max}\left( \mathbf{U}_{p}^{\left( \tau _2 \right)} \right) \geqslant \epsilon ^{\left( \tau _2 \right)}\mathrm{Tr}\left( \mathbf{U}_p \right) 
	, \label{PB_relaxed:b} \\
	&   \ \ \ \ \  \eqref{OP2:b},~\eqref{ABO_1:b} ,~\eqref{PB:b}-\eqref{PB:f}. \label{PB_relaxed:c} 	
	\end{align}
\end{subequations} 	

 Problem~\eqref{PB_relaxed} is a standard convex SDP, which can be solved efficiently by numerical solvers such as the SDP solver in CVX tool~\cite{cvx}. The parameter $\epsilon ^{\left( \tau _2 \right)}$ can be updated via~\cite{9139273,9423667}
   \begin{equation} \label{update epsilon}
 	\epsilon ^{\left( \tau _2+1 \right)}=\min \left( 1,\frac{\varepsilon _{\max}\left( \mathbf{U}_{p}^{\left( \tau _2 \right)} \right)}{\mathrm{Tr}\left( \mathbf{U}_{p}^{\left( \tau _2 \right)} \right)}+\varDelta ^{\left( \tau _2 \right)} \right), 
 \end{equation} 
where $\varDelta ^{\left( \tau _2 \right)}$ is the step size. 	
 	
The details of the proposed sequential constraint relaxation algorithm is presented in Algorithm~\ref{Passive algorithm}. For the convergence analysis of Algorithm~\ref{Passive algorithm}, similar proof can be found in~\cite{sequential}, 	
 \begin{algorithm} 
	\caption{Sequential constraint relaxation
		Algorithm for obtaining $\left\{ \mathbf{U}_{p}^{*} \right\} $}
	\label{Passive algorithm}
	\begin{algorithmic}[1]
		\STATE  Initialize feasible points $\mathbf{U}_{p}^{\left( 0 \right)}$, step size $\varDelta ^{\left( 0 \right)}$, error tolerance $\varrho $ , set relaxation parameter $\epsilon ^{\left( \tau _2 \right)}=0$ and the iteration index ${\tau_2} = 0$.
		\REPEAT
		\STATE  Solve problem~\eqref{PB_relaxed} to obtain $\mathbf{U}_{p}$;
		\STATE \textbf{if} problem~\eqref{PB_relaxed} is solvable 
		\STATE ~~~Update $\mathbf{U}_{p}^{\left( \tau_2+1 \right)}=\mathbf{U}_{p}$; 
		\STATE ~~~Update $\varDelta ^{\left( \tau _2+1 \right)}=\varDelta ^{\left( 0 \right)}$;
		\STATE \textbf{else}  
		\STATE ~~~Update $\mathbf{U}_{p}^{\left( \tau_2+1 \right)}=\mathbf{U}_{p}^{\left( \tau_2 \right)}$;  
		\STATE ~~~update $\varDelta ^{\left( \tau _2+1 \right)}=\frac{\varDelta ^{\left( \tau _2 \right)}}{2}
		$;
		\STATE \textbf{end}
		\STATE  Update ${\tau_2} = {\tau_2} + 1$;
		\STATE  Update $\epsilon ^{\left( \tau _2+1 \right)}$ via~\eqref{update epsilon};		
		\UNTIL {$\left| 1-\epsilon ^{\left( \tau _2 \right)} \right|\leqslant \varrho $ and the objective value of problem~\eqref{PB_relaxed} converge.}
		\STATE   \textbf{Output}: $\mathbf{U}_{p}^{*}$
	\end{algorithmic}
\end{algorithm} 	
\subsection{Proposed Algorithm, Convergence and Complexity}  	
Based on the above discussions, we provide the details of the proposed two-layer iterative algorithm to solve the original problem~\eqref{OP1} in {Algorithm~\ref{overall Algorithm}}. Specifically, the inner-layer iteration is to solve the joint optimization problem over power allocation coefficients, active beamforming vectors, transmission and reflection beamforming vectors by alternating optimization. The outer-layer iteration is mainly to update the decoding order with the solutions obtained from the inner-layer iteration. 
\begin{algorithm}
	\caption{Two-Layer Iterative Algorithm }
	\label{overall Algorithm}
	\begin{algorithmic}[1]
		\STATE  Initialize $\left\{ \rho _{c,k}^{\left( 0 \right)} \right\} $, $\left\{  \mathbf{w}_{c}^{\left( 0 \right)} \right\}$, $\left\{ \mathbf{u}_{p}^{\left( 0 \right)} \right\}$ and error tolerance $\varDelta$;  Set the iteration index ${\tau_0} = 0$.
		\REPEAT
		\STATE  Calculate the equivalent-combined channel gains $\left\{ \Gamma _{c,\mathcal{D} _c\left( k \right)} \right\} $ via~\eqref{equivalent channel gain};
		\STATE  Update the decoding order $\left\{ \mathcal{D} _c \right\} $ based on Lemma~\ref{decoding order};
	    \STATE  Calculate the achievable sum rate $R_{\mathrm{sum}}\left( \rho _{c,k}^{\left( \tau_0 \right)},\mathbf{w}_{c}^{\left( \tau_0 \right)},\mathbf{u}_{p}^{\left( \tau_0 \right)} \right)  $;
	         	 ~~ \REPEAT 
		\STATE   Update the power allocation factors $\left\{ \rho _{c,k} \right\} $ according to Theorem~\ref{Theorem optimal power};		
		\STATE   Update active beamforming vectors $\left\{ \textbf{w} _{c} \right\} $ via Algorithm~\ref{Active algorithm};
		\STATE   Update passive beamforming vectors $\left\{ \textbf{u} _{p} \right\} $ via Algorithm~\ref{Passive algorithm};
   				 \UNTIL the objective value of problem~\eqref{OP2} converges. 
   	    \STATE   Record the obtained solutions $\left\{ \rho _{c,k} \right\} $, $\left\{ \textbf{w} _{c} \right\} $, $\left\{ \textbf{u} _{p} \right\} $; 
   	    \STATE   calculate the achievable sum rate $R_{\mathrm{sum}}\left( \rho _{c,k},\mathbf{w}_c,\mathbf{u}_p \right)$;
   	    \STATE \textbf{if} $R_{\rm sum}\left( \rho _{c,k},\mathbf{w}_c,\mathbf{u}_p \right) \geqslant R_{\rm sum}\left( \rho _{c,k}^{\left( \tau_0 \right)},\mathbf{w}_{c}^{\left( \tau_0 \right)},\mathbf{u}_{p}^{\left( \tau_0 \right)} \right)$  
   	    \STATE   ~~ $\rho _{c,k}^{\left( \tau_0+1 \right)}=\rho _{c,k},\mathbf{w}_{c}^{\left( \tau_0+1 \right)}=\mathbf{w}_c,\mathbf{u}_{p}^{\left( \tau_0+1 \right)}=\mathbf{u}_p,  k\in \mathcal{K} _c, c\in \mathcal{C} , p=\left\{ t,r \right\}$;
   	    \STATE \textbf{else}
   	    	\STATE   ~~ $\rho _{c,k}^{\left( \tau_0+1 \right)}=\rho _{c,k}^{\left( \tau_0 \right)},\mathbf{w}_{c}^{\left( \tau_0+1 \right)}=\mathbf{w}_{c}^{\left( \tau_0 \right)},\mathbf{u}_{p}^{\left( \tau_0+1 \right)}=\mathbf{u}_{p}^{\left( \tau_0 \right)},  k\in \mathcal{K} _c, c\in \mathcal{C} , p=\left\{ t,r \right\} $;
   	    \STATE \textbf{end if}	
   	    \STATE update $\tau_0=\tau_0+1$;
		\UNTIL $\begin{array}{l} \frac{\left| R_{\rm sum}\left( \rho _{c,k}^{\left( \tau_0 \right)},\mathbf{w}_{c}^{\left( \tau_0 \right)},\mathbf{u}_{p}^{\left( \tau_0 \right)} \right) -R_{\rm sum}\left( \rho _{c,k}^{\left( \tau_0-1 \right)},\mathbf{w}_{c}^{\left( \tau_0-1 \right)},\mathbf{u}_{p}^{\left( \tau_0-1 \right)} \right) \right|}{R_{\rm sum}\left( \rho _{c,k}^{\left( t \right)},\mathbf{w}_{c}^{\left( \tau_0 \right)},\mathbf{u}_{p}^{\left( \tau_0 \right)} \right)} \\ <\varDelta \end{array}$
		\STATE   \textbf{Output}: the optimal $\left\{ \rho _{c,k}^* \right\} $, $\left\{ \textbf{w} _{c}^* \right\} $, $\left\{ \textbf{u} _{p}^* \right\} $;
	\end{algorithmic}
\end{algorithm} 
\subsubsection{Convergence analysis}
To begin with, we first prove the convergence of the inner-layer iteration. For a given decoding order, the power allocation coefficients $\left\{ \rho _{c,k} \right\} $, active beamforming vectors $\left\{ \mathbf{w}_c \right\} $, transmission and reflection beamforming vectors $\left\{ \mathbf{u}_p \right\} $ in problem~\eqref{OP2} are solved alternatingly, we have the following inequality: 
\begin{equation}\label{converge proof}
\begin{split}
&R_{\mathrm{sum}}\left( \rho _{c,k}^{\left( \tau_3 \right)},\mathbf{w}_{c}^{\left( \tau_3 \right)},\mathbf{u}_{p}^{\left( \tau_3 \right)} \right) \\
& \overset{\left( a \right)}{\leqslant}R_{\mathrm{sum}}\left( \rho _{c,k}^{\left( \tau_3+1 \right)},\mathbf{w}_{c}^{\left( \tau_3 \right)},\mathbf{u}_{p}^{\left( \tau_3 \right)} \right) 
\\
& \overset{\left( b \right)}{\leqslant}R_{\mathrm{sum}}\left( \rho _{c,k}^{\left(\tau_3+1 \right)},\mathbf{w}_{c}^{\left( \tau_3+1 \right)},\mathbf{u}_{p}^{\left( \tau_3 \right)} \right) 
\\
& \overset{\left( c \right)}{\leqslant}R_{\mathrm{sum}}\left( \rho _{c,k}^{\left(\tau_3+1 \right)},\mathbf{w}_{c}^{\left( \tau_3+1 \right)},\mathbf{u}_{p}^{\left( \tau_3+1 \right)} \right) ,
\end{split} 
\end{equation} 
where (a) holds since for fixed $\left\{ \mathbf{w}_{c}^{\left( \tau_3 \right)},\mathbf{u}_{p}^{\left( \tau_3 \right)} \right\} $, the optimal $\left\{ \rho _{c,k}^{\left(\tau_3+1 \right)} \right\} $ is obtained according to Theorem~\ref{Theorem optimal power}; (b) comes from that the solution $\left\{ \mathbf{w}_{c}^{\left(\tau_3+1 \right)} \right\} $ is solved via Algorithm 1 with $\left\{ \rho _{c,k}^{\left( \tau_3+1 \right)},\mathbf{u}_{p}^{\left( \tau_3\right)} \right\} 
$; (c) follows the fact that the optimal $\left\{ \mathbf{u}_{p}^{\left( \tau_3+1 \right)} \right\} $ is obtained by Algorithm 2 with given $\left\{ \rho _{c,k}^{\left( \tau_3+1 \right)},\mathbf{w}_{c}^{\left(\tau_3 +1 \right)} \right\} $. $\tau_3$ is the inner-layer iteration index.
 
The inequalities in~\eqref{converge proof} indicates that the objective value of problem~\eqref{OP2} is monotonically non-decreasing after each iteration. On the other hand, the achievable sum rate is upper bounded.
Therefore, the inner-layer iteration is guaranteed to converge. For the outer-layer iteration, it is easy observed from step 13 to 17 that the achievable sum rate is monotonically non-decreasing after each iteration. Since both the inner- and outer-layer iterations converge, the proposed Algorithm 3 converges. 
\subsubsection{Complexity analysis}
It is observed that the complexity of Algorithm~\ref{overall Algorithm} mainly depends on that of Algorithms~\ref{Active algorithm} and~\ref{Passive algorithm}. Therefore, we only analyses the complexity of the two algorithms. The complexity of Algorithm 1 to solve the active beamforming optimization problem~\eqref{OP3} is $O_1\triangleq \mathcal{O} \left( \tau _{1}^{\max}\max \left( N_{\mathrm{T}},\left( 2K+1 \right) \right) ^4\sqrt{N_{\mathrm{T}}}\log _2\frac{1}{\epsilon _1} \right) 
$, where $\tau _{1}^{\max}$ is the number of iterations for Algorithm~\ref{Active algorithm} and $\epsilon _1$ is the solution accuracy. The complexity of Algorithm~\ref{Passive algorithm} to solve the transmission and reflection beamforming optimization problem~\eqref{PB} is $O_2\triangleq \mathcal{O} \left( \tau _{2}^{\max}\max \left( M,\left( 2K \right) \right) ^4\sqrt{M}\log _2\frac{1}{\epsilon _2} \right) $, where $\tau _{2}^{\max}$ is the number of iterations for Algorithm~\ref{Passive algorithm} and $\epsilon _2$ is the solution accuracy. As a result, the total complexity of Algorithm~\ref{overall Algorithm} is $\mathcal{O} \left( \tau _{0}^{\max}\tau _{3}^{\max}\left( O_1+O_2 \right) \right) $ where $\tau _{0}^{\max}$ and $\tau _{3}^{\max}$ are the iteration numbers of the outer- and inner-layer iterations in Algorithm~\ref{overall Algorithm}.
 \section{NUMERICAL RESULTS}
\begin{figure}[t!]
	\centering
	\includegraphics[scale=0.13]{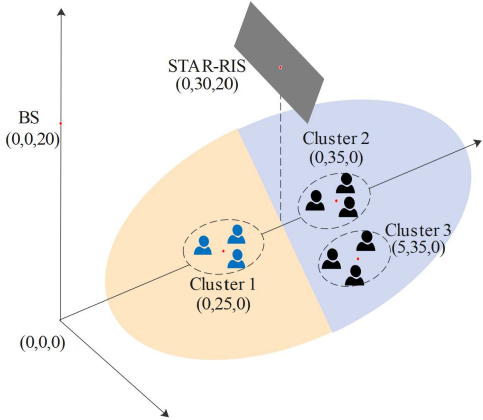}
	\caption{Simulation scenario of the STAR-RIS-NOMA system}
	\label{simulation_scenario}
\end{figure}

In this section, numerical simulations are conducted to evaluate the performance of the proposed algorithm. Without loss of generality, we assume that there are three clusters in the STAR-RIS-NOMA system and each cluster contains three users. The considered simulation scenario is shown in Fig.~\ref{simulation_scenario}. Specifically, cluster 1 is located in the reflection space, and clusters 2 and 3 are located in the transmission space. \textcolor[rgb]{0.00,0.00,0.00}{The BS and STAR-RIS are located at $\left( 0,~0,~20\right) $ meters and $\left( 0,~30,~20\right) $ meters,  respectively. The central coordinates of the three clusters are set as $\left( 0,~25,~0\right) $ meters, $\left( 0,~35,~0\right) $ meters, and $\left( 5,~30,~0\right) $ meters, respectively. The users are randomly located in their corresponding clusters with a radius of $5~\rm m$.} The distance-dependent channel path loss is modeled as $\mathcal{P} \left( d \right) =\varepsilon _0\left( \frac{d}{d_0} \right) ^{-\ell}$, where $\varepsilon _0$ is the path loss at the reference distance $d_0=1$ meter (m), $d$ denotes the link distance and $\ell$ denotes the path loss exponent. To model small-scale fading,  we adopt \textcolor[rgb]{0.00,0.00,0.00}{Rician} fading for all channels involved. For example, the Rician fading of the BS-RIS channel $\textbf{F}$ is given by $\textbf{F}_{\rm Ric}=\sqrt{\frac{\kappa_{\rm BR}}{1+\kappa_{\rm BR}}}\textbf{F}_{\rm Ric}^{\text{LoS}}+\sqrt{\frac{1}{1+\kappa_{\rm BR}}}\textbf{F}_{\rm Ric}^{\text{NLoS}}$, where $\kappa_{\rm BR}$ is the \textcolor[rgb]{0.00,0.00,0.00}{Rician} factor, ${{{\textbf{F}}}_{\rm Ric}^{\rm Los}}$ and ${{{\textbf{F}}}_{\rm Ric}^{\rm NLoS}}$ are the line-of-sight (LoS) component and non-LoS (NLoS) component, respectively. The RIS-user channels $\left\{ \mathbf{g}_{c,k} \right\} $ can be similarly modeled according to the above model. \textcolor[rgb]{0.00,0.00,0.00}{The path loss and path loss exponent of the BS-RIS link are denoted by $\varepsilon ^{\rm BR}$ and $\ell^{\mathrm{BR}}$, respectively. Similarly, $\varepsilon^{\mathrm{RU}}_{c,k}$, $\ell^{\mathrm{RU}}_{c,k}$ and $\kappa _{c,k}^{\mathrm{RU}}$ denotes the path loss, path loss exponent and Rician factor for the channel from the RIS to user $k$ in cluster $c$, respectively. Without loss of generality, we assume that the users have the same QoS requirements and we set $R_{c,k}^{\min}=0.1~\rm bits/s/Hz$. For the BS-RIS and RIS-user links, the path loss exponents are assumed to be the same and set to $\ell^{\mathrm{BR}}=\ell^{\mathrm{RU}}_{c,k}=2.2$~\cite{9197675}. Similarly, the path loss are set to be $\varepsilon ^{\rm BR} =\varepsilon^{\mathrm{RU}}_{c,k} =-30~\rm dB$~\cite{Zhang2021SecuringNN,9570143} and the Rician factors are set to $\kappa_{\rm BR}=\kappa _{c,k}^{\mathrm{RU}}=3~\rm dB$~\cite{9353406}. The noise power is set to $\sigma ^2=-90~\rm dBm$~\cite{9570143}. The adopted parameter values are summarized in Table II unless otherwise specified.}
\begin{table*}[t!]
 	\caption{Simulation Parameters}
 	\label{parameter}
 	\begin{center}
 		\begin{tabular} {|l|l|}
 			\hline
 				Parameter &  Value \\
 			\hline
 			 	The locations of the BS and RIS & $\left( 0,~0,~20\right) $, $\left( 0,~30,~20\right) $		\\	
 			\hline
 				The central coordinates of the three clusters' circle  &  $\left( 0,~25,~0\right) $, $\left( 0,~35,~0\right) $, $\left( 5,~30,~0\right) $\\
 			\hline
 				The radius of the three groups' circle & $ 5$~{\rm m}, $5$~{\rm m}, $5$~{\rm m}   \\
			\hline 			 				
				The path loss exponents of the BS-RIS and RIS-user links	       &  $ 2.2, 2.2$   \\ 	
			\hline
			    The path loss at 1 meter      &		$-30$~{\rm dB}	 \\			
 			\hline
 			    The Rician factors of the BS-RIS and RIS-user links	       &    $3$~{\rm dB}, $3$~{\rm dB}     \\
 		     \hline 	
 			     The minimum QoS requirement for each user & $0.1$ bits/s/Hz \\
        	\hline
 				Noise power     &    \textcolor[rgb]{0.00,0.00,0.00}{$-90$~{\rm dBm}} \\ 			
 			\hline 			
 		\end{tabular}
 	\end{center}
 \end{table*}
 \subsection{Convergence of The Proposed Algorithms}
\begin{figure*}[t!]
 	\centering
 	\subfigure[Convergence behavior of Algorithm 1, $P_{\max}=35 {\rm dBm}$, $M=10$]
 	{
 		\begin{minipage}[t]{0.45\linewidth}
 			\label{Algorithm1_convergence}
 			\includegraphics[scale=0.5]{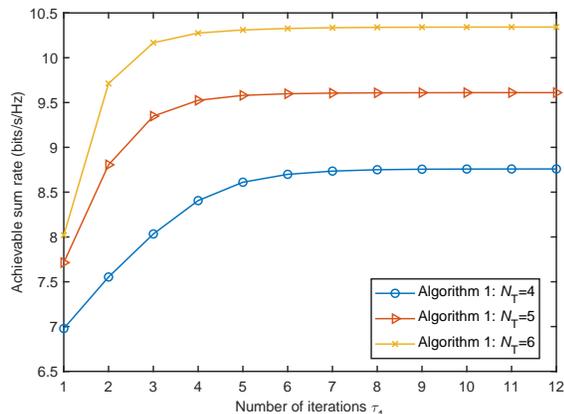}
 		\end{minipage}%
 	}%
  	\subfigure[Convergence behavior of Algorithm 2, $P_{\max}=35 {\rm dBm}$, $N_{\rm T}=4$]
  	{
 	\begin{minipage}[t]{0.45\linewidth}
 		\label{Algorithm2_convergence}
 		\includegraphics[scale=0.5]{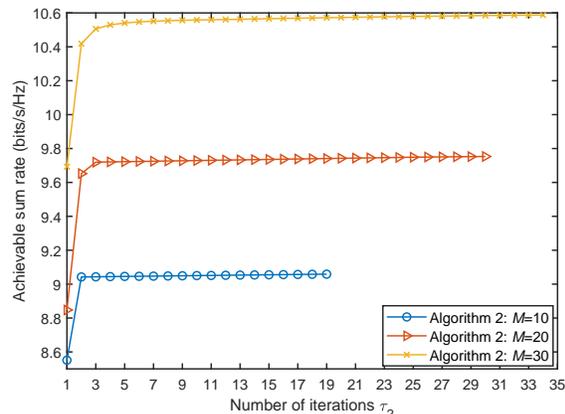}
 	\end{minipage}%
 	}%
\\
  	\subfigure[Convergence behavior of inner-layer iteration of Algorithm 3, $N_{\rm T}=4$, $M=10$]
  	{
	\begin{minipage}[t]{0.45\linewidth}
		\label{inner_converge}
		\includegraphics[scale=0.5]{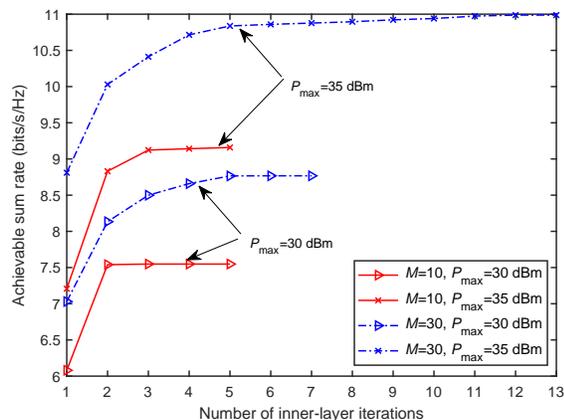}
	\end{minipage}%
	}%
  	\subfigure[Convergence behavior of outer-layer of Algorithm 3, $N_{\rm T}=4$, $M=10$]
  	{
	\begin{minipage}[t]{0.45\linewidth}	  
	\label{outer_converge}
		\includegraphics[scale=0.5]{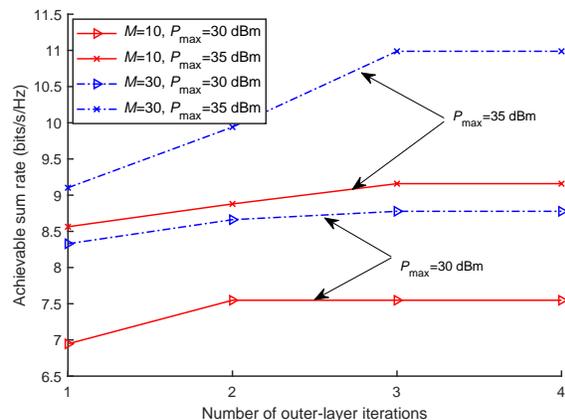}
	\end{minipage}%
	}%
\caption{Convergence behavior of the proposed algorithms}
\label{convergence}
 \end{figure*}

 In Figs.~\ref{convergence}, the convergence of the proposed algorithms are analyzed by using numerical simulations. Specifically, Fig.~\ref{Algorithm1_convergence} depicts the convergence of Algorithm 1 against the number of iterations $\tau_1$. It is observed that the active beamforming optimization algorithm converges within 12 iterations under different
 settings of the number of BS antennas. Fig.~\ref{Algorithm2_convergence} investigates the convergence of Algorithm 2. From Fig.~\ref{Algorithm2_convergence}, it can be found that the number of iterations for the convergence of
 Algorithm 2 increases with $M$, because more transmission and reflection coefficients need to be optimized. 
 
 \textcolor[rgb]{0.00,0.00,0.00}{Fig.~\ref{inner_converge} depicts the convergence of the inner-layer iteration under different settings of $M$ and $P_{\max}$. It is shown that the number of iterations for the convergence increases with $M$ and $P_{\max}$ increase. This is because, $P_{\max}$ is the total transmit power budget of the active beamforming vectors, larger value of $P_{\max}$ leads to larger feasible solution region of the active beamforming vectors. In addition, more transmission and reflection coefficients need to be optimized as $M$ increases. The outer-layer iteration is designed to update the decoding order. In the worst case, the iteration number for the convergence of the outer-layer iteration is equal to the total number of all possible decoding order. To show the effectiveness of the proposed outer-layer algorithm, Fig.~\ref{outer_converge} evaluates the convergence under different settings of $M$ and $P_{\max}$. As can be seen, the outer-layer iteration can converge quickly within a small number of iterations, which confirms the effectiveness of our proposed algorithms.}
\subsection{Impact of The Number of RIS Elements}
\begin{figure}[!t]
 \setlength{\belowcaptionskip}{-0.55cm}   
	\centering
	\includegraphics[scale=0.55]{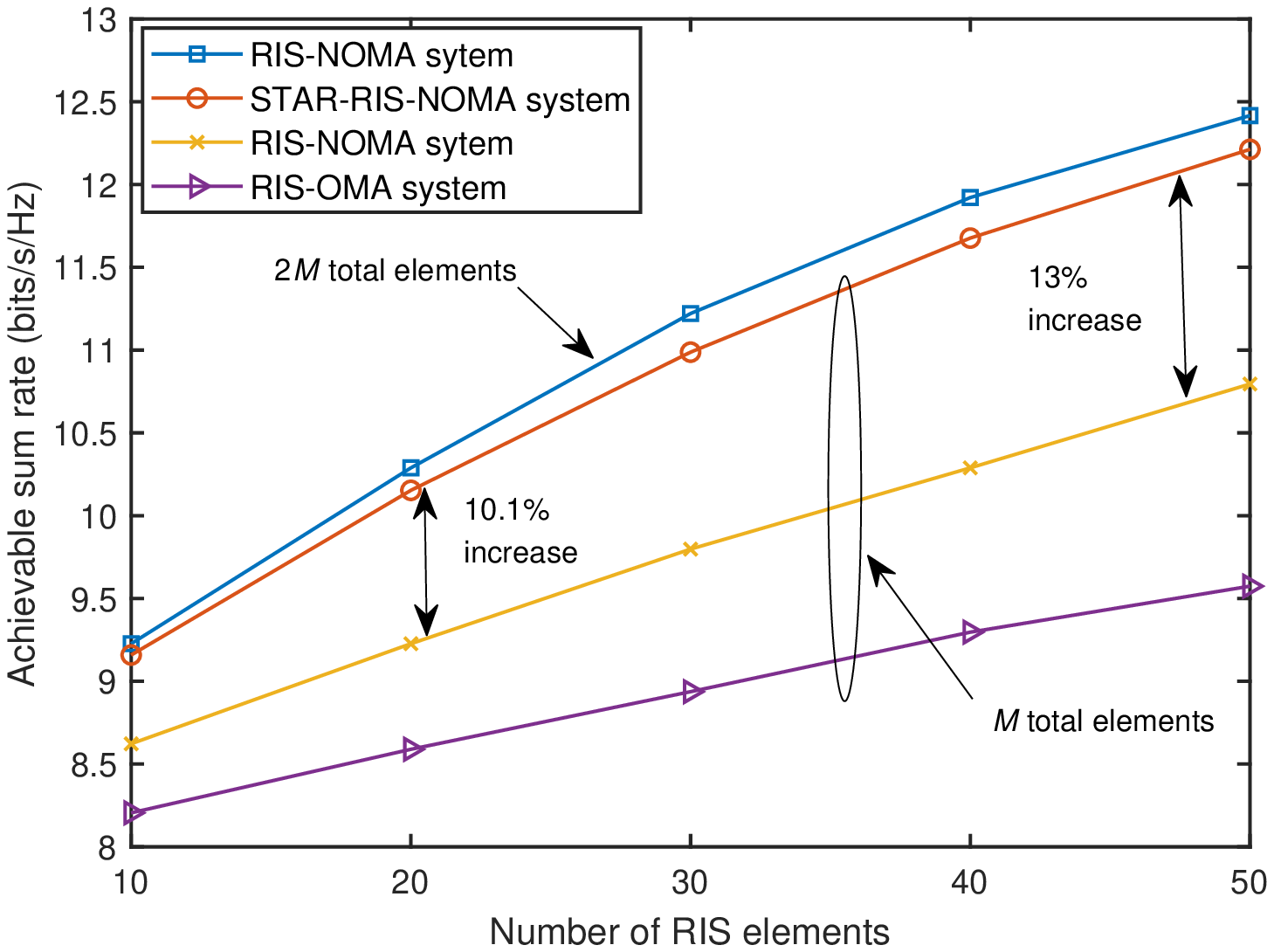}
	\caption{The achievable sum rate versus the number of RIS elements, $N_{\rm T}=4$}
	\label{Rsum_M}
 \end{figure}

To evaluate the performance of the proposed algorithm for the STAR-RIS-NOMA system, two benchmark schemes are considered, namely, traditional RIS-NOMA and traditional RIS-OMA systems. For the two traditional RIS assisted systems, to achieve the full-space coverage, one transmitting-only RIS and one reflecting-only RIS are employed and deployed adjacent to each other at the same location as the STAR-RIS. For the RIS-OMA system, the BS serves all users through time
division multiple access with the aid of the traditional reflecting/transmitting-only RISs. 
\textcolor[rgb]{0.00,0.00,0.00}{Furthermore, we consider two cases that each traditional reflecting/transmitting-only RISs contains $M/2$ and $M$ elements, respectively. Thus, the number of total elements corresponding to the above two cases are $M$ and $2M$.} Fig.~\ref{Rsum_M} shows the achievable sum rate versus the number of RIS elements $M$. \textcolor[rgb]{0.00,0.00,0.00}{It is first observed that the STAR-RIS-NOMA system outperforms RIS-NOMA system with $M$ total elements, because that the STAR-RIS-NOMA system has $M$ elements in both transmission and reflection beamforming compared to $M/2$ elements for RIS-NOMA systems. In addition, the performance gap between STAR-RIS-NOMA and RIS-NOMA system increases as $M$ increases, since the RIS can not flexibly choose between transmission and reflection modes, the DoFs will significantly loss. Secondly, the RIS-NOMA system with $2M$ total elements outperforms the STAR-RIS-NOMA system. This can be explained as follows. The total elements in both transmission and reflection beamforming are the same. For STAR-RIS-NOMA system, the energy of the signal incident on each STAR-RIS elements is split into two parts. However, for the RIS-NOMA system, the energy of the signal on each side of the RIS elements keeps unchanged. As a result, the received signal strength of RIS-NOMA system with $2M$ total elements is stronger than that of the STAR-RIS-NOMA system.}  Third, the NOMA based systems outperform the OMA based system. This is expected since users can be served simultaneously through the NOMA protocol compared with the OMA scheme.
 
\begin{figure}[!t] 
	\centering
	\includegraphics[scale=0.55]{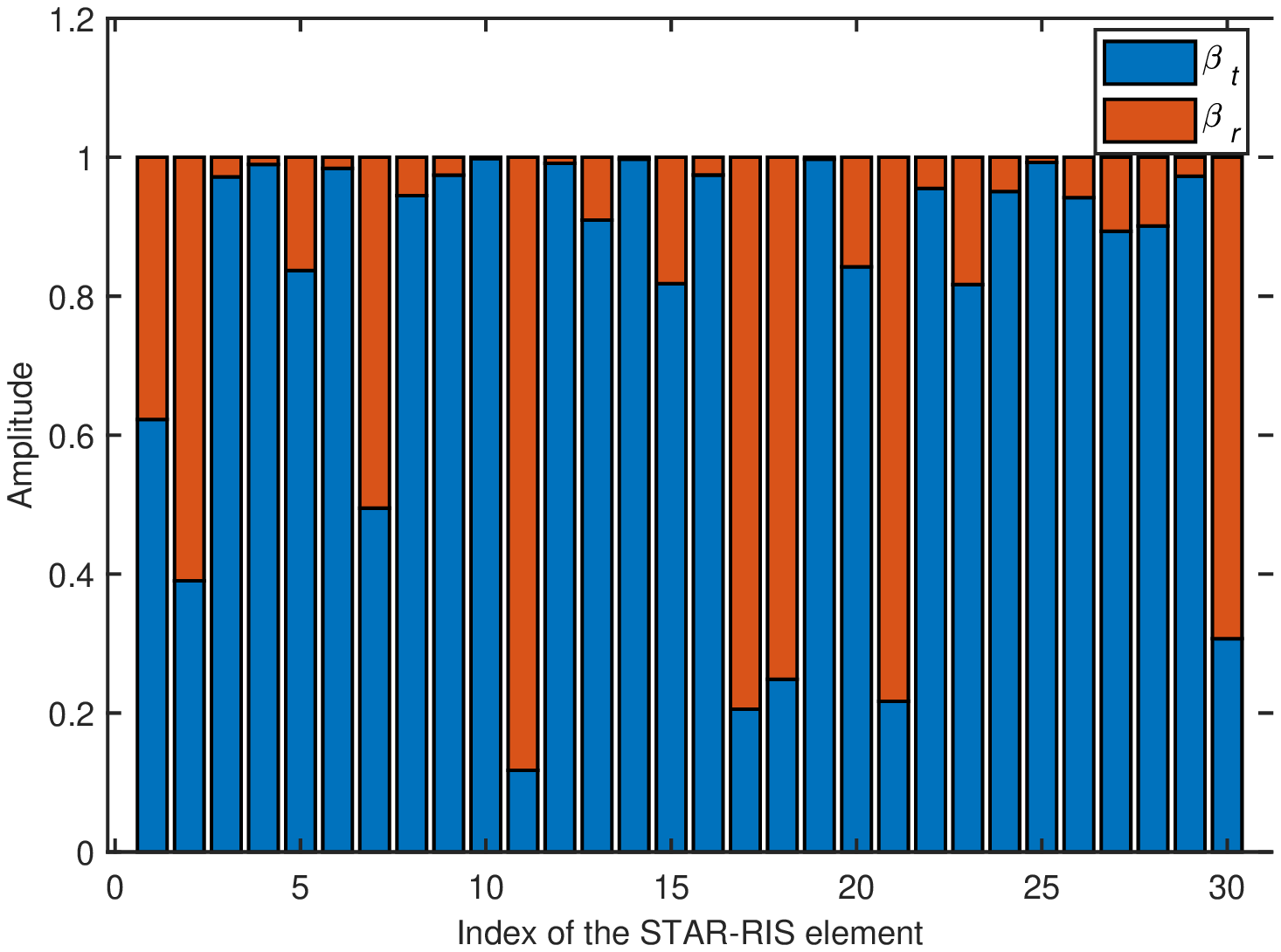}
	\caption{The value of the transmission and reflection amplitudes for each STAR-RIS element, $M=30$, $N_{\rm T}=4$, $P_{\rm max}=35~\rm {dBm}$}
	\label{Amplitude}
\end{figure}

Fig.~\ref{Amplitude} plots the value of the transmission and reflection amplitudes for each STAR-RIS element. From Fig.~\ref{Amplitude}, we can observe that the energy allocated to the transmission amplitudes $\left\{ \beta _t \right\} $ are more than that allocated to the reflection amplitudes $\left\{ \beta _r \right\}$. This phenomenon can be explained as that there are two clusters in the transmission space, and one cluster in the reflection space, to enhance the received signal power for each user, the STAR-RIS allocates more energy to the transmission amplitudes. 
 \subsection{Impact of The number of BS antennas}
\begin{figure}[!t]
   \setlength{\belowcaptionskip}{-0.55cm}   
 	\centering
	\includegraphics[scale=0.55]{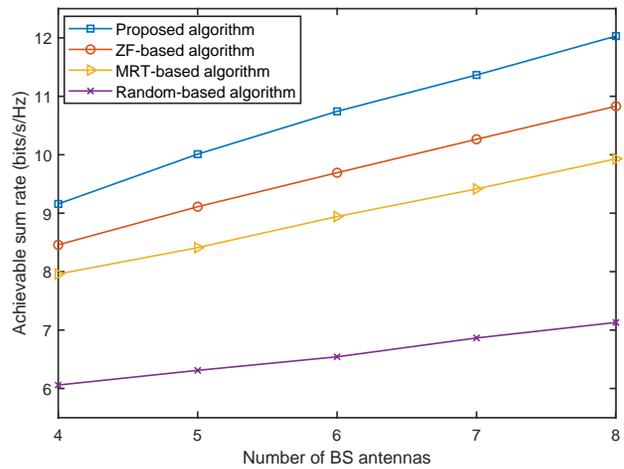}
	\caption{The achievable sum rate versus the number of BS antennas, $M=10$, $P_{\rm max}=35 \rm {dBm}$}
\label{Rsum_NT}
 \end{figure}

 Fig.~\ref{Rsum_NT} presents the achievable sum rate of the proposed STAR-RIS-NOMA system versus the number of BS antennas $N_{\rm T}$. For comparison, the performance of three benchmark schemes, namely, ZF-based algorithm, MRT-based algorithm and Random-based algorithm are considered as well. Specifically, the active beamforming vectors of the ZF- and Random-based algorithms are obtained by zero-forcing (ZF) method~\cite{9174801} and random selection method. For the MRT-based algorithm, the active beamforming vectors are solved by the maximum-ratio transmission (MRT) method~\cite{8811733}, which aims to maximize the combined channel gain of the user with the highest decoding order in each cluster. In addition, the power allocation coefficients, transmission and reflection beamforming vectors for the above schemes are solved by our proposed algorithm. As it can be seen from Fig.~\ref{Rsum_NT}, the achievable sum rate of all schemes increases as $N_{\rm T}$ increases. This is expected since larger $N_{\rm T}$ enable a higher active beamforming gain. In addition, a considerable performance loss is observed from the proposed algorithm and the benchmark algorithms, which confirms the importance of joint optimization over the vectors of active, transmission and reflection beamforming. 
\subsection{Impact of Decoding Order}
\begin{figure}[!t] 
	\centering
	\includegraphics[scale=0.55]{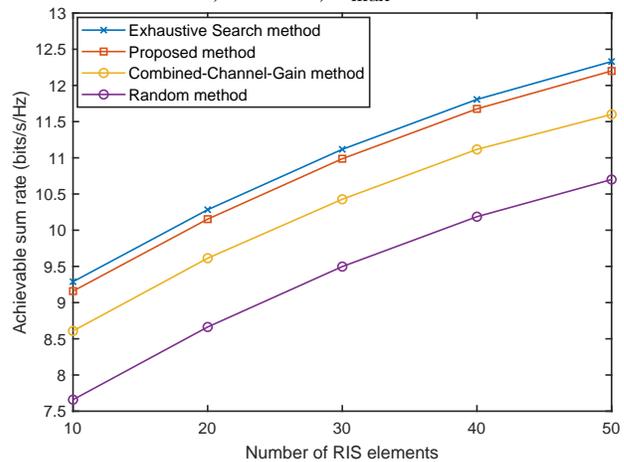}
	\caption{The achievable sum rate versus the number of RIS elements, $N_{\rm T}=4$, $P_{\rm max}=35 \rm {dBm}$}
	\label{Decoding_order_M}
  \end{figure}

In Fig.~\ref{Decoding_order_M}, we evaluate the impact of the decoding order on the
achievable sum rate performance. Three decoding order determination methods are compared with our proposed method. The first benchmark scheme named \textcolor[rgb]{0.00,0.00,0.00}{Exhaustive Search method}, finds the optimal decoding order via exhaustive search. The second  benchmark scheme named Random method, selects the decoding order randomly. The third benchmark scheme named Combined-Channel-Gain method, employs combined channel gains to determine the decoding order. In particular, for Combined-Channel-Gain method, the decoding order in each cluster is determined as follows:
 \begin{equation}
 \begin{split}
   &\left| \mathbf{g}_{c,\mathcal{D} _c\left( 1 \right)}^{H}\mathbf{\Theta }_{c,\mathcal{D} _c\left( 1 \right)}\mathbf{Fw}_c \right|^2\leqslant \left| \mathbf{g}_{c,\mathcal{D} _c\left( 2 \right)}^{H}\mathbf{\Theta }_{c,\mathcal{D} _c\left( 2 \right)}\mathbf{Fw}_c \right|^2\leqslant \cdots \\ 
   & \leqslant \left| \mathbf{g}_{c,\mathcal{D} _c\left( K_c \right)}^{H}\mathbf{\Theta }_{c,\mathcal{D} _c\left( K_c \right)}\mathbf{Fw}_c \right|^2.
   \end{split}
 \end{equation}
 
   It is noted that the above decoding order depends on the vectors of active, transmission and reflection beamforming, which can also be updated by our proposed algorithms. From Fig.~\ref{Decoding_order_M}, it can be found that the  proposed method can achieve performance close to that achieved by the \textcolor[rgb]{0.00,0.00,0.00}{Exhaustive Search method}. Though some performance loss is incurred by the proposed method, the complexity of the proposed method is much lower than that of the \textcolor[rgb]{0.00,0.00,0.00}{Exhaustive Search method}. In addition, our proposed method outperforms the Combined-Channel-Gain and Random methods. The reasons behind this can be explained as follows. For single-cluster RIS-NOMA systems, since there is no inter-cluster interference, the optimal decoding order is determined by the users' combined channel gains. However, for the multiple-cluster RIS-NOMA systems, the decoding order is determined not only by the users' combined channel gains, but also by the interference channel gains from other clusters. Therefore, according to Theorem~\ref{Theorem optimal power}, our proposed decoding order determination method is more reasonable. 
   \subsection{Impact of The Total Transmit Power Budget}
\begin{figure}[!t]
     \setlength{\abovecaptionskip}{3pt}
   \setlength{\belowcaptionskip}{-20pt}
  	\centering
  	\includegraphics[scale=0.55]{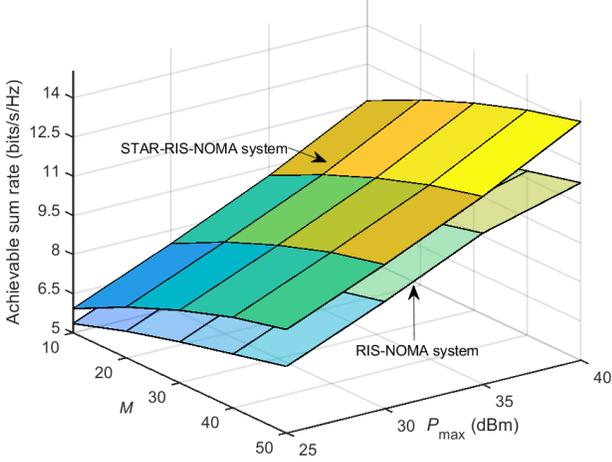}
  	\caption{The achievable sum rate versus the total transmit power budget under different number of RIS elements, $N_{\rm T}=4$}
  	\label{Rsum_M_PT}
  \end{figure}

  In Fig.~\ref{Rsum_M_PT}, the impact of the total transmit power budget under different number of RIS elements are analyzed. It can be found that the proposed STAR-RIS-NOMA system is always capable of outperforming the traditional RIS-NOMA system \textcolor[rgb]{0.00,0.00,0.00}{with $M$ total elements,} which implies the effectiveness of the proposed system again. In addition, for fixed number of RIS elements $M$, the achievable sum rate of both systems increases as the total transmit power budget $P_{\rm max}$ increases.   
    \section{Conclusions} 
 In this paper, we have investigated a new joint optimization problem for STAR-RIS-NOMA systems, where the decoding order, power allocation coefficients, active beamforming, transmission and reflection beamforming are jointly optimized to  maximize the achievable sum rate. We have proposed a novel two-layer iterative algorithm to solve the formulated non-convex optimization problem. In particular, the joint optimization problem over power allocation coefficients, active beamforming, transmission and reflection beamforming is solved by an inner-layer iteration. With the obtained solutions, the decoding order in each cluster is then updated by an outer-layer iteration. In addition, an efficient decoding order determination scheme and closed-form power allocation strategy has been proposed. Simulation results have validated the effectiveness of the proposed STAR-RIS-NOMA system. Moreover, it has been found that the proposed decoding order determination scheme can achieve near-optimal performance. 
 
 \textcolor[rgb]{0.00,0.00,0.00}{In this work, we assume that the phase-shift coefficients for transmission and reflection can be independently adjusted. However, this assumption may be too ideal for purely passive STAR-RISs. In the real application scenarios, the phase-shift coefficients for
 	transmission and reflection are coupled with each other, which will not only cause performance degradation but also make the transmission and reflection beamforming design much more complicated. Therefore, the joint design for STAR-RIS with coupled phase-shift is an important topic to investigate in future work. }
  \vspace{-0.4cm}
 \section*{Appendix~A: Proof of Proposition~\ref{proposition of lemma 1}} \label{corollary prove}
\renewcommand{\theequation}{A.\arabic{equation}}
\setcounter{equation}{0}  
 
 For any users $k$ and $j$ in cluster $c$ with the optimal decoding order $\mathcal{D} _{c}^{-1}\left( j \right) >\mathcal{D} _{c}^{-1}\left( k \right)$, according to Lemma~\ref{decoding order}, the equivalent-combined channel gains of the two users must satisfy the following condition
 \begin{equation}\label{corollary 1}
  \setlength{\abovedisplayskip}{2pt}
  \setlength{\belowdisplayskip}{2pt}
 	\frac{\left| \mathbf{g}_{c,j}^{H}\mathbf{\Theta }_{c,j}\mathbf{Fw}_c \right|^2}{\sum_{\underline{c}\ne c}{\left| \mathbf{g}_{c,j}^{H}\mathbf{\Theta }_{c,j}\mathbf{Fw}_{\underline{c}} \right|^2}+\sigma ^2}\geqslant \frac{\left| \mathbf{g}_{c,k}^{H}\mathbf{\Theta }_{c,k}\mathbf{Fw}_c \right|^2}{\sum_{\underline{c}\ne c}{\left| \mathbf{g}_{c,k}^{H}\mathbf{\Theta }_{c,k}\mathbf{Fw}_{\underline{c}} \right|^2}+\sigma ^2}.
 \end{equation}
 
 \eqref{corollary 1} can be further rewritten as 
 \begin{equation}\label{corollary 1-1}
 \begin{split}
 \setlength{\abovedisplayskip}{2pt}
 \setlength{\belowdisplayskip}{2pt}
  & \left| \mathbf{g}_{c,j}^{H}\mathbf{\Theta }_{c,j}\mathbf{Fw}_c \right|^2\left( \sum_{\underline{c}\ne c}{\left| \mathbf{g}_{c,k}^{H}\mathbf{\Theta }_{c,k}\mathbf{Fw}_{\underline{c}} \right|^2}+\sigma ^2 \right)  \\
  & \geqslant \left| \mathbf{g}_{c,k}^{H}\mathbf{\Theta }_{c,k}\mathbf{Fw}_c \right|^2\left( \sum_{\underline{c}\ne c}{\left| \mathbf{g}_{c,j}^{H}\mathbf{\Theta }_{c,j}\mathbf{Fw}_{\underline{c}} \right|^2}+\sigma ^2 \right). 
 \end{split}
 \end{equation}
	
We first multiply both sides of~\eqref{corollary 1-1} by $\rho _{c,k}$, and then add $\rho _{c,k}\left| \mathbf{g}_{c,j}^{H}\mathbf{\Theta }_{c,j}\mathbf{Fw}_c \right|^2\left| \mathbf{g}_{c,k}^{H}\mathbf{\Theta }_{c,k}\mathbf{Fw}_c \right|^2\sum_{n>k}{\rho _{c,n}}
$ to both sides. As a result, we can obtain the inequality as in Equation~\eqref{corollary 2} at the top of next page. 
 \begin{figure*}[!t]
	\normalsize
\begin{equation}\label{corollary 2}
 \setlength{\abovedisplayskip}{2pt}
 \setlength{\belowdisplayskip}{2pt}
  \begin{split}
	& \left| \mathbf{g}_{c,j}^{H}\mathbf{\Theta }_{c,j}\mathbf{Fw}_c \right|^2\rho _{c,k}\left( \left| \mathbf{g}_{c,k}^{H}\mathbf{\Theta }_{c,k}\mathbf{Fw}_c \right|^2\sum_{n\in \mathcal{K}_c,n>k}{\rho _{c,n}}+\sum_{\underline{c}\in \mathcal{C},\underline{c}\ne c}{\left| \mathbf{g}_{c,k}^{H}\mathbf{\Theta }_{c,k}\mathbf{Fw}_{\underline{c}} \right|^2}+\sigma ^2 \right) 
	\\
	&\geqslant \left| \mathbf{g}_{c,k}^{H}\mathbf{\Theta }_{c,k}\mathbf{Fw}_c \right|^2\rho _{c,k}\left( \left| \mathbf{g}_{c,j}^{H}\mathbf{\Theta }_{c,j}\mathbf{Fw}_c \right|^2\sum_{n\in \mathcal{K}_c,n>k}{\rho _{c,n}}+\sum_{\underline{c}\in \mathcal{C},\underline{c}\ne c}{\left| \mathbf{g}_{c,j}^{H}\mathbf{\Theta }_{c,j}\mathbf{Fw}_{\underline{c}} \right|^2}+\sigma ^2 \right) .
  \end{split}	
\end{equation}
 	\hrulefill \vspace*{0pt}
\end{figure*} 

By rearranging~\eqref{corollary 2}, we have:
\begin{equation}\label{corollary 3}
  \mathrm{SINR}_{j\rightarrow k}^{c}\geqslant \mathrm{SINR}_{k\rightarrow k}^{c},
\end{equation}
which means that the SIC condition $R_{j\rightarrow k}^{c}\geqslant R_{k\rightarrow k}^{c}$ is guaranteed.
 \section*{Appendix~B: Proof of Lemma~\ref{feasible}} \label{feasible prove}
\renewcommand{\theequation}{B.\arabic{equation}}
\setcounter{equation}{0} 
  From the SINR~\eqref{SINRkk} and the equivalent-combined channel gain~\eqref{equivalent channel gain}, the achievable data rate $R_{k\rightarrow k}^{m}$ can be rewritten as  
 \begin{equation}\label{rewritten Rkk}
 	R_{k\rightarrow k}^{c}=\log _2\left( 1+\frac{\Gamma _{c,k}\rho _{c,k}}{\Gamma _{c,k}\sum_{n\in \mathcal{K}_c,n>k}{\rho _{c,n}}+1} \right), 
 \end{equation}
 
Let $\rho _{c,k}^{\min}$ denote the minimum power allocation coefficient for user $k$ in cluster $c$. If the users in any cluster are allocated their minimum power allocation factor, then all the users will achieve their minimum QoS requirement. Thus, we have the following equation:
\begin{equation} \label{Rmin equation}
	\log _2\left( 1+\frac{\Gamma _{c,k}\rho _{c,k}^{\min}}{\Gamma _{c,k}\sum_{n\in \mathcal{K}_c,n>k}{\rho _{c,n}^{\min}}+1} \right) =R_{c,k}^{\min}.
 \end{equation}	
		
Then, the minimum power allocation coefficient is given by
\begin{equation} \label{min power} 
	\rho _{c,k}^{\min}=r_{c,k}^{\min}\left( \sum_{n\in \mathcal{K}_c,n>k}{\rho _{c,n}^{\min}}+\frac{1}{\Gamma _{c,k}} \right), 
 \end{equation} 
 where $r_{c,k}^{\min}=2^{R_{c,k}^{\min}}-1$.
 
 As a result, we can obtain the minimum sum power allocation factor $\rho _{c}^{\min}$ in cluster $c$ 
\begin{equation} \label{min sum power} 	
	\rho _{c}^{\min}=\sum_{k=1}^{K_c}{\rho _{c,k}^{\min}}=\sum_{k=1}^{K_c}{\frac{r_{c,k}^{\min}}{\Gamma _{c,k}}\prod_{n=1}^{k-1}{\left( r_{c,n}^{\min}+1 \right)}},
  \end{equation} 	
where $\prod_{i=1}^{0}{\left( r_{c,n}^{\min}+1 \right)}=1$. 	
 	
Finally, Lemma~\ref{feasible} is proved. 
 \section*{Appendix~C: Proof of Theorem~\ref{Theorem optimal power}} 
\renewcommand{\theequation}{C.\arabic{equation}}
\setcounter{equation}{0}	 	
 Following the idea introduced in~\cite{8454272,7973036,8672163}, the user with the higher decoding order are allocated as much power as possible if the other users meet the minimum rate requirements. Specifically, for any cluster $c$, the power allocation should be firstly performed for the users with weaker equivalent combined channel gains and only the minimum power should be allocated to maintain the minimum QoS requirements. In addition, it is optimal to allocate all of the remaining power to the user with the strongest equivalent combined channel gain. Thus, we have the following equation set:
 \begin{equation} \label{optimal power equation set} 
  \setlength{\belowdisplayskip}{2pt}
 \left\{ \begin{array}{c}
 	R_{k\rightarrow k}^{c}=R_{c,k}^{\min},k=1,2,\cdots ,K_c-1,\\
 	\sum_{k=1}^{K_c}{\rho _{c,k}}=1.\\
 \end{array} \right. 
  \end{equation}  	
 
 The equation set~\eqref{optimal power equation set} can be further expressed as:
  \begin{equation} \label{optimal power equation set1}
 \begin{cases}
 	\rho _{c,1}=r_{c,1}^{\min}\left( 1-\rho _{c,1}+\frac{1}{\Gamma _{c,1}} \right),\\
 \rho _{c,2}=r_{c,1}^{\min}\left( 1-\sum_{n=1}^2{\rho _{c,n}}+\frac{1}{\Gamma _{c,2}} \right),\\
 \vdots\\
 \rho _{c,K_c-1}=r_{c,K_c-1}^{\min}\left( 1-\sum_{n=1}^{K_c-1}{\rho _{c,n}}+\frac{1}{\Gamma _{c,K_c-1}} \right),\\
 \sum_{k=1}^{K_c}{\rho _{c,k}}=1.
 \end{cases}
 \end{equation}
 	
Then, by solving the above equation set, we obtain the optimal power allocation factors $\left\{ \rho _{c,k}^{*} \right\} $ of problem~\eqref{power allocation}, which is given by~\eqref{optimal power}. Since $R_{k\rightarrow k}^{c}=R_{c,k}^{\min}\left( k=1,2,\cdots ,K_c-1 \right) 
$ and $\rho _{c,K_c}^{*}=1-\sum_{i=1}^{K_c-1}{\rho _{c,i}^{*}}$, the optimal objective value of problem~\eqref{optimal power} can be expressed as
\begin{equation}\label{max sum rate in prove}
  \setlength{\abovedisplayskip}{2pt}
 \setlength{\belowdisplayskip}{2pt}
R_{\mathrm{sum}}=\sum_{c=1}^C{\sum_{k=2}^{K_c-1}{R_{c,k}^{\min}}}+\sum_{c=1}^C{\log _2\left( 1+\rho _{c,K_c}^{*}\Gamma _{c,K_c} \right)}.
\end{equation} 	
\vspace{-0.5cm}
 \section*{Appendix~D: Proof of Theorem~\ref{rank 1}} \label{Proof rank1}
 \renewcommand{\theequation}{D.\arabic{equation}}
 \setcounter{equation}{0}	
Problem~\eqref{ABO_1} without the rank-one constraint is jointly convex with respect to all optimization variables. Hence, the optimal solution to problem~\eqref{ABO_1} can be obtained by its corresponding dual problem with zero duality gap. Specifically, the Lagrangian function of problem~\eqref{ABO_1} in terms of beamforming matrix $\mathbf{W}_{c}^{*}$ is given as in Equation~\eqref{Lagrangian} at the top of next page, 
 \begin{figure*}[!t]
	\normalsize
 \begin{equation} \label{Lagrangian}  
  \setlength{\abovedisplayskip}{2pt}
 \setlength{\belowdisplayskip}{2pt}
   \begin{split}
		\mathcal{L} = & \sum_{c\in \mathcal{C}}{\sum_{k\in \mathcal{K} _c}{\beta _{c,k}\mathrm{Tr}\left( \mathbf{W}_c\mathbf{H}_{c,k} \right) \rho _{c,k}}}-\lambda \sum_{c\in \mathcal{C}}{\mathrm{Tr}\left( \mathbf{W}_c \right)}+\sum_{c\in \mathcal{C}}{\mathrm{Tr}\left( \mathbf{X}_c\mathbf{W}_c \right)}
		\\
		&-\sum_{c\in \mathcal{C}}{\sum_{k\in \mathcal{K} _c}{\chi _{c,k}\left( -\mathrm{Tr}\left( \mathbf{W}_c\mathbf{H}_{c,k} \right) \sum_{n\in \mathcal{K}_c,n>k}{\rho _{c,n}}+\sum_{\underline{c}\in \mathcal{C},\underline{c}\ne c}{\mathrm{Tr}\left( \mathbf{W}_{\underline{c}}\mathbf{H}_{c,k} \right)} \right)}}+\Gamma ,
 	\end{split} 
 \end{equation}
  	\hrulefill \vspace*{0pt}
\end{figure*}  
where $\Gamma$ denotes the collection of terms that only involve
variables that are not relevant for the proof, $\beta _{c,k}$, $\chi _{c,k}$ and $\lambda$ are the Lagrange multiplier corresponding to constraint~\eqref{ABO:b},~\eqref{ABO:c} and~\eqref{ABO:d}, matrix $\mathbf{X}_c$ is the Lagrange multiplier matrix for the positive semi-definite constraint on matrix $\mathbf{W}_c$ in constraint~\eqref{ABO:f}. 
 
The dual problem of~\eqref{ABO} is given by:
 \begin{equation} \label{dual}
	\underset{\beta _{c,k},\chi _{c,k},\lambda \geqslant 0,\mathbf{X}_c\succcurlyeq 0}{\max}\underset{\mathbf{W}_c,A_{c,k},B_{c,k}}{\min}\mathcal{L} .
 \end{equation}

The KKT conditions for the optimal $\mathbf{W}_{c}^{*}$ are given by:
\begin{numcases}{}
 \setlength{\abovedisplayskip}{2pt}
 \setlength{\belowdisplayskip}{2pt}
	\beta _{c,k}^{*},\chi _{c,k}^{*},\lambda ^*\geqslant 0, \label{KKT1}\\
	\mathbf{X}_{c}^{*}\succcurlyeq 0, \label{KKT2}\\
	\mathbf{X}_{c}^{*}\mathbf{W}_{c}^{*}=\mathbf{0}, \label{KKT3}\\
	\frac{\partial \mathcal{L}}{\partial \mathbf{W}_c}=\mathbf{0}\Longrightarrow \mathbf{X}_{c}^{*}=\lambda \mathbf{I}-\mathbf{Q}_{c}^{*}, \label{KKT4} 
\end{numcases}
where $\beta _{c,k}^{*}$, $\chi _{c,k}^{*}$, $\lambda ^*$ are the optimal Lagrange multipliers. It is noted that there exists at least one $\lambda ^*>0$, since constraint~\eqref{ABO:d} is active for optimal $\mathbf{W}_{c}^{*}$. The matrix $\mathbf{Q}_{c}^{*}$ is defined as
 \begin{equation} \label{matrix Q}
 \begin{split}
 \setlength{\abovedisplayskip}{2pt}
  \setlength{\belowdisplayskip}{2pt}
	\mathbf{Q}_{c}^{*}= & \sum_{k\in \mathcal{K} _c}{\left( \beta _{c,k}^{*}\rho _{c,k}^{*}-\chi _{c,k}^{*}\sum_{n\in \mathcal{K}_c,n>k}{\rho _{c,n}} \right) \mathbf{H}_{c,k}^{H}} \\
	&-\sum_{\underline{c}\in \mathcal{C},\underline{c}\ne c}{\sum_{k\in \mathcal{K} _c}{\chi _{\underline{c},k}^{*}\mathbf{H}_{\underline{c},k}^{H}}}.
	 \end{split}
 \end{equation}

 Noted that $\mathbf{X}_{c}^{*}$ is a positive semidefinite matrix, according to the results in~\cite{9183907}, we have $\mathrm{rank}\left( \mathbf{X}_{c}^{*} \right) =N_{\mathrm{T}}-1$. In addition, considering the constraint in~\eqref{KKT3}, we have:$\mathrm{rank}\left( \mathbf{W}_{c}^{*} \right) +\,\,\mathrm{rank}\left( \mathbf{X}_{c}^{*} \right) \leqslant N_{\mathrm{T}}$. Moreover, we can conclude that $\mathrm{rank}\left( \mathbf{W}_{c}^{*} \right) =1$, because that $\mathrm{rank}\left( \mathbf{W}_{c}^{*} \right) =0$, i.e., $\mathbf{W}_{c}^{*}=\textbf{0}$ is contradict with the minimum QoS requirement constraints of problem~\eqref{ABO_1}.

\vspace{-0.5cm}
\bibliographystyle{IEEEtran}
\bibliography{zjkbib}

\end{document}